\documentclass[aps,prl,reprint,groupedaddress,showpacs,longbibliography]{revtex4-1}

\usepackage{siunitx}
\usepackage{amssymb,amsmath} 
\usepackage{graphicx}
\usepackage{xcolor}
\usepackage{pgfplots}
\usepgfplotslibrary{groupplots}

\newcommand{\micron}{$\si{\micro m}$ }
\newcommand{\microsec}{$\si{\micro s}$}
\usepackage{pgfplots}

\usepackage{stackrel}
\usepackage[normalem]{ulem}

\begin{document}

\title{Magnetic dipolar interaction between hyperfine clock states in a planar alkali Bose gas}

\author{Y.-Q. Zou}
\author{B. Bakkali-Hassani}
\author{C. Maury}
\author{\'E. Le Cerf}
\author{S. Nascimbene}
\author{J. Dalibard}
\author{J. Beugnon}

\email[]{beugnon@lkb.ens.fr}

\affiliation{Laboratoire Kastler Brossel,  Coll\`ege de France, CNRS, ENS-PSL University, Sorbonne Universit\'e, 11 Place Marcelin Berthelot, 75005 Paris, France}

\date{\today}

\begin{abstract}
In atomic systems, clock states feature a zero projection of the total angular momentum and thus a low sensitivity to magnetic fields. This makes them widely used for metrological applications like atomic fountains or gravimeters. Here, we show that a mixture of two such non-magnetic states still display magnetic dipole-dipole interactions comparable to the one expected for the other Zeeman states of the same atomic species. Using high resolution spectroscopy of a planar gas of $^{87}$Rb atoms with a controlled in-plane shape, we explore the effective isotropic and extensive character of these interactions and demonstrate their tunability. Our measurements set strong constraints on the relative values of the $s$-wave scattering lengths $a_{ij}$ involving the two clock states.
\end{abstract}

\maketitle

Quantum atomic gases constitute unique systems to investigate many-body physics thanks to the precision with which one can control their interactions \cite{Chin08,Bloch08}. Usually, in the ultra-low temperature regime achieved with these gases, contact interactions described by the $s$-wave scattering length dominate. In recent years, non-local interaction potentials have been added to the quantum gas toolbox. Long-range interactions can be mediated thanks to optical cavities inside which atoms are trapped \cite{Baumann10}. Electric dipole-dipole interactions are routinely achieved via excitation of atoms in Rydberg electronic states \cite{Low12}. Atomic species with large magnetic moments in the ground state, like Cr, Er or Dy, offer the possibility to explore the role of magnetic dipole-dipole interactions (MDDI) \cite{Lahaye09}. The latter case has led, for instance, to the observation of quantum droplets \cite{Ferrier16}, roton modes \cite{Chomaz18}, or spin dynamics in lattices with off-site interactions \cite{Paz13,Baier16,lepoutre2019out}. 

\begin{figure}[t!]
  \begin{flushleft}
    \hskip0pt\textbf{(a)}
    \hskip100pt\textbf{(b)}
  \end{flushleft}  
  \includegraphics[width=0.22\textwidth]{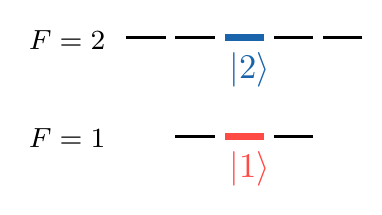} 
  \includegraphics[width=0.22\textwidth]{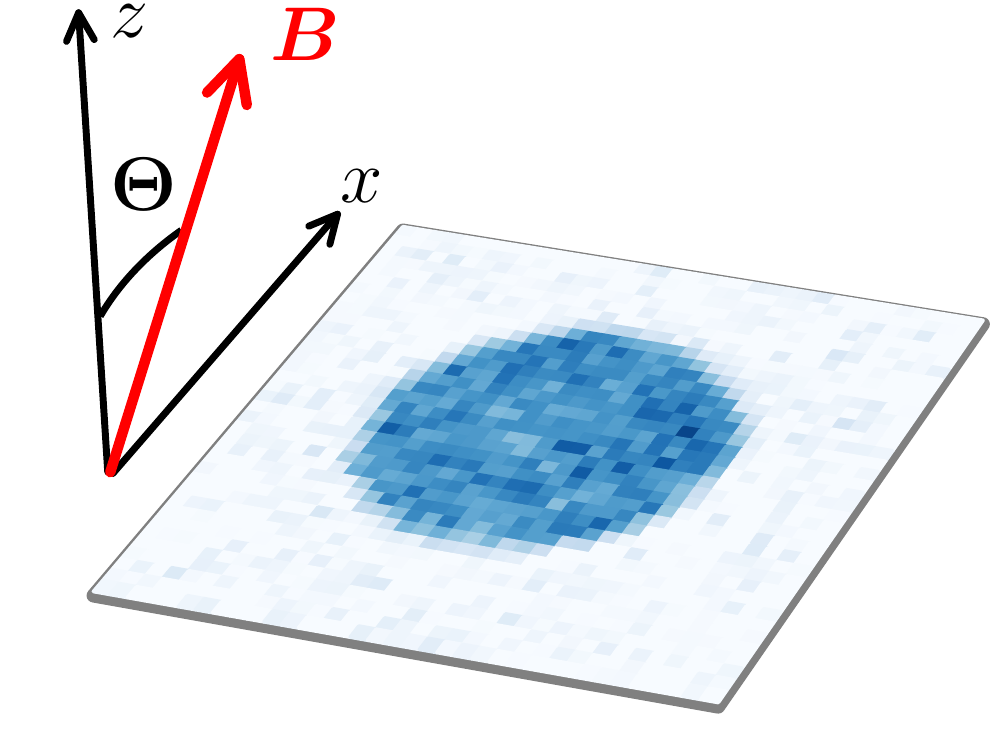}
  \vskip 10mm
  \includegraphics[width=75mm]{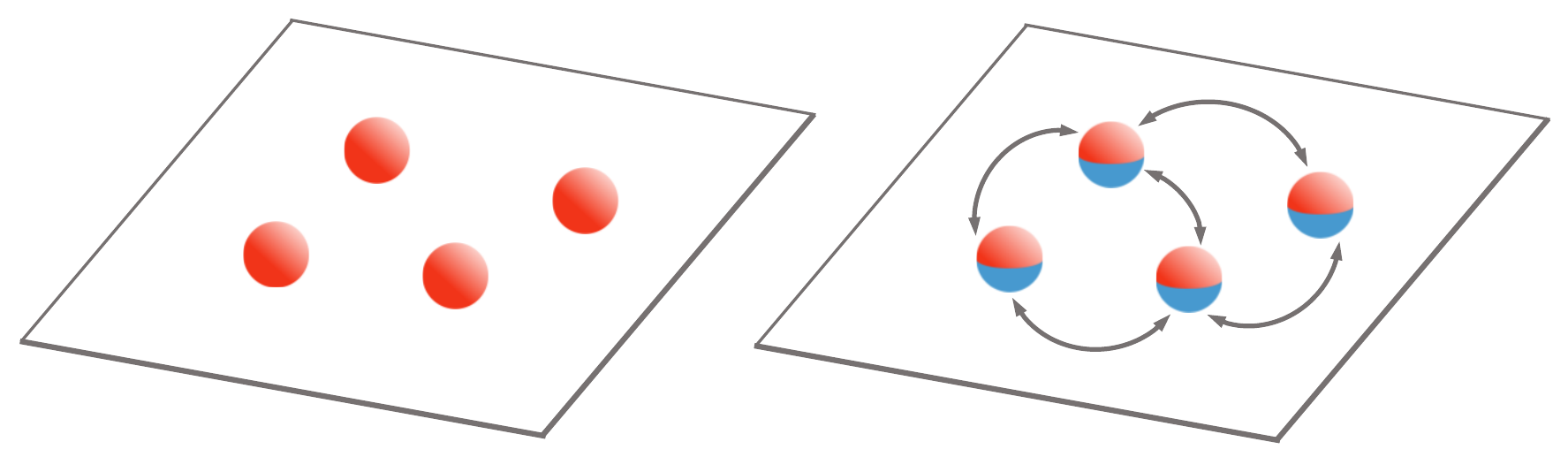}
  \begin{flushleft}
       \vskip-30mm\textbf{(c)}\hskip 38mm\textbf{(d)}\vskip30mm
  \end{flushleft}
  \vskip-20pt
  \caption{(a)  Level diagram of the hyperfine ground-level manifold showing the two states relevant to this work  $|1\rangle\equiv|F=1, m_Z=0\rangle$ and $|2\rangle \equiv|F=2, m_Z=0\rangle$. (b)  Image of the atomic cloud obtained through absorption imaging along the direction perpendicular to the atomic plane. Atoms are confined in the $xy$ plane in a disk of radius 12\,$\micron$. The orientation of the magnetic field  ${\boldsymbol B}$ is tuned in the $xz$ plane. (c) Schematics of atoms prepared in the state $|1\rangle$, with no MDDI in this case. MDDI are also absent when all atoms are in $|2\rangle$. (d) Significant MDDI occur for atoms in a linear superposition of $|1\rangle$ and $|2\rangle$.}
   \label{fig1}
\end{figure}

For alkali-metal atoms, which are the workhorse of many cold-atom experiments, the magnetic moment is limited to $\lesssim 1$\,Bohr magneton ($\mu_B$) and in most cases, MDDI have no sizeable effect on the gas properties \cite{Giovanazzi02}. However, some paths have been investigated to evidence their role also for these atomic species. A first route consists in specifically nulling the $s$-wave scattering length using a Feshbach resonance \cite{Fattori08,Pollack09}, so that MDDI become dominant. A second possibility is to operate with a multi-component (or spinor) gas \cite{StamperKurn13}, using several states from the ground-level manifold of the atoms. One can then take advantage of a possible coincidence of the various scattering lengths in play. When it occurs, the spin-dependent contact interaction is much weaker than the spin-independent one, and MDDI can have a significant effect  \cite{Yi04}, e.g.  on the generation of spin textures \cite{Vengalattore08, Eto14} and on magnon spectra \cite{Marti14}.  In all instances studied so far with these multi-component gases, each component possesses a non-zero magnetic moment and creates a magnetic field that influences its own dynamics, as well as the dynamics of the other component(s). 

In this Letter, we present another, yet unexplored, context in which MDDI can influence significantly the physics of a two-component gas of alkali-metal atoms. We operate with a superposition of the two hyperfine states of $^{87}$Rb involved in the so-called hyperfine clock transition, $|1\rangle\equiv|F=1, m_Z=0\rangle$ and $|2\rangle \equiv|F=2, m_Z=0\rangle$, where the quantization axis $Z$ is aligned with the uniform external magnetic field (Fig.\,\ref{fig1}a). For a single-component gas prepared in one of these two states, the average magnetization is zero by symmetry and MDDI have no effect. However, when atoms are simultaneously present in these two states, we show that magnetic interactions between them are non-zero, and that the corresponding MDDI can modify significantly the position of the clock transition frequency.

Our work constitutes a magnetic analog of the observation of electric dipole-dipole interactions (EDDI) between molecules in a Ramsey interferometric scheme \cite{yan2013observation}. There, in spite of the null value of the electric dipole moment of a molecule prepared in an energy eigenstate, it was shown that EDDI can be induced in a molecular gas by preparing a coherent superposition of two rotational states. Both in our work and in \cite{yan2013observation}, the coupling between two partners results in a pure exchange interaction, with one partner switching from $|1\rangle$ to $|2\rangle$, and the other one from $|2\rangle$ to $|1\rangle$. This exchange Hamiltonian also appears for resonant EDDI between atoms prepared in different Rydberg states \cite{Leseleuc:2017_PhysRevLett.119.053202}. 
        
In spite of their different origin, the physical manifestations of MDDI in our setup are similar to the standard ones. Here, we study it for a 2D gas using high-resolution Ramsey spectroscopy (Fig.\,\ref{fig1}b) and we explicitly test two important features of DDI in this planar geometry: their effect does not depend on the in-plane shape of the cloud (isotropy), nor on its size (extensivity). More precisely, we recast the role of MDDI as a modification of the $s$-wave inter-species scattering length $a_{12}$, and show the continuous tuning of $a_{12}$ by changing the orientation of the external magnetic field with respect to the atom plane. We obtain in this way an accurate information on the relative values of intra- and inter-species bare scattering lengths of the studied states. 

We start with the restriction of the MDDI Hamiltonian to the clock state manifold \footnote{For all experiments reported here, the fraction of atoms in any other spin state remains below our detection sensitivity of 1\%.}, using the magnetic interaction between two electronic spins $\hat {\boldsymbol{s}}_A$ and $\hat {\boldsymbol{s}}_B$ with magnetic moments $\boldsymbol{m}_{A,B}=2\mu_B \boldsymbol{s}_{A,B}$ \begin{eqnarray}
\hat V_{\rm dd}(r,\boldsymbol{u})=\frac{\mu_0 \mu_B^2}{\pi r^3 } [\hat {\boldsymbol{s}}_A \cdot \hat {\boldsymbol{s}}_B-3 (\hat {\boldsymbol{s}}_A \cdot \boldsymbol{u})(\hat {\boldsymbol{s}}_B  \cdot \boldsymbol{u})],\label{eq:VDD}
\end{eqnarray}
where $r$ is the distance between the two dipoles and $\boldsymbol{u}$ is the unit vector connecting them. The calculation detailed in 
\cite{REFSM} shows that MDDI do not modify the interactions between atoms in the same state $|1\rangle$ or $|2\rangle$, but induce a non-local, angle-dependent, exchange interaction (Figs.\,\ref{fig1}cd). The second-quantized Hamiltonian of the MDDI for the clock states is thus:
\begin{align}
\hat H_{\rm dd}^{(1,2)}&=\frac{\mu_0 \mu_B^2}{4\pi }\iint \mathrm{d}^3r_A\; \mathrm{d}^3r_B  \nonumber\\ 
&\frac{1-3 \cos^2\theta}{r^3}\  \hat \Psi_2^\dagger(\boldsymbol{r}_A)\, \hat \Psi_1^\dagger(\boldsymbol{r}_B)\, \hat \Psi_2(\boldsymbol{r}_B)\, \hat \Psi_1(\boldsymbol{r}_A),
\label{eq:Hamiltonian}
\end{align}
where the $\hat \Psi_i(\boldsymbol{r}_\alpha)$ are the field operators annihilating a particle in state $|i\rangle$ at position $\boldsymbol{r}_\alpha$, $r=|\boldsymbol r_A-\boldsymbol r_B|$ and  $\theta$ is the angle between  $\boldsymbol r_A-\boldsymbol r_B$ and the quantization axis.

We now investigate the spatial average value of $\hat H_{\rm dd}^{(1,2)}$. We note first that for a 3D isotropic gas, the angular integration gives $\langle \hat H_{\textrm {dd}}^{(1,2)}\rangle_{\textrm{3D}}=0$, as usual for MDDI \cite{Lahaye09}. 
We then consider a homogeneous quasi-2D Bose gas confined isotropically in the $xy$ plane with area $L^2$. We assume that the gas has a Gaussian density profile along the third direction $z$,  $n_{1,2}(z)=N_{1,2} e^{-z^2/\ell_z^2}/\sqrt{\pi} \ell_z L^2$, where $\ell_z=\sqrt{\hbar/m \omega_z}$ is the extension of the ground state of the harmonic confinement of frequency $\omega_z$ for particles of mass $m$ and $N_{1,2}$ is the atom number in states $|1\rangle$, $|2\rangle$. One then finds \cite{Fischer06,Fedorov14,Mishra16}:
\begin{eqnarray}
\langle \hat H_{\textrm {dd}}^{(1,2)}\rangle_{\textrm{2D}}=\frac{ \mu_0 \mu_B^2 N_1 N_2}{3 \sqrt{2\pi} \ell_z L^2 } (3 \cos^2\Theta -1),\label{eq:DDMF2D}
\end{eqnarray} 
where $\Theta$ is the angle between the external magnetic field $\boldsymbol{B}$ and the direction perpendicular to the atomic plane. This energy is maximal and positive for $\boldsymbol{B}$ perpendicular to the atomic plane ($\Theta=0$), and minimal and negative for $\boldsymbol{B}$ in the atomic plane ($\Theta=\pi/2$).  Eq.\,(\ref{eq:DDMF2D}) shows that the energy per atom in state $|1\rangle$ depends only on the spatial density $N_2/L^2$ of atoms in state $|2\rangle$, which proves the extensivity. 

In 2D, the Fourier transform of the dipole-dipole Hamiltonian possesses a well-defined value at the origin $\boldsymbol{k}=0$ \cite{Fischer06}. Consequently, for a large enough sample (typically $L\gg \ell_z$), the average energy $\langle \hat H_{\textrm {dd}}^{(1,2)}\rangle_{\textrm{2D}}$, evaluated by switching the integral (\ref{eq:Hamiltonian}) to Fourier space, is independent of the system shape. This contrasts with the 3D case, for which the MDDI energy changes sign when switching from an oblate to a prolate cloud \cite{Yi00,Lahaye09}. Considering the effective isotropy of the MDDI in this 2D configuration, it is convenient to describe their role as a change $\delta a_{12}$ of the inter-species scattering length with respect to its bare value defined as  $a_{12}^{(0)}$. In 2D, interspecies contact interactions lead to $\langle\hat H_{\textrm {contact}}^{(1,2)}\rangle_{\textrm{2D}}=\sqrt{8 \pi }\, a_{12}\,\hbar^2 N_1N_2/(m \ell_z L^2) $ and   we deduce
\begin{eqnarray}
\delta a_{12}(\Theta)=a_{\rm dd}\left(3 \cos^2\Theta -1\right),
\label{eq:da12}
\end{eqnarray}
where $a_{\rm dd}=\mu_0 \mu_B^2 m /(12 \pi \hbar^2)$ is the so-called dipole length that quantifies the strength of MDDI \footnote{We use the definition of Ref. \cite{Lahaye09}. Other definitions with a different numerical factor are found in the literature, see for instance, \cite{Bortolotti06,Baranov12}}. 

\begin{figure}[t]
  \begin{flushleft}
       \vskip10pt\textbf{(a)}\hskip125pt\textbf{(b)}
  \end{flushleft}
  \vskip-20pt
  \includegraphics{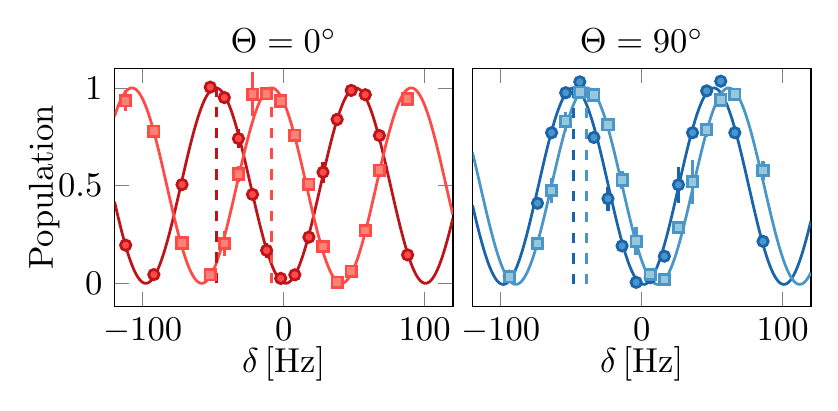}\\
  \begin{flushleft}
      \vskip-20pt 
      \textbf{(c)}
  \end{flushleft}
  \vskip-5mm
  \includegraphics{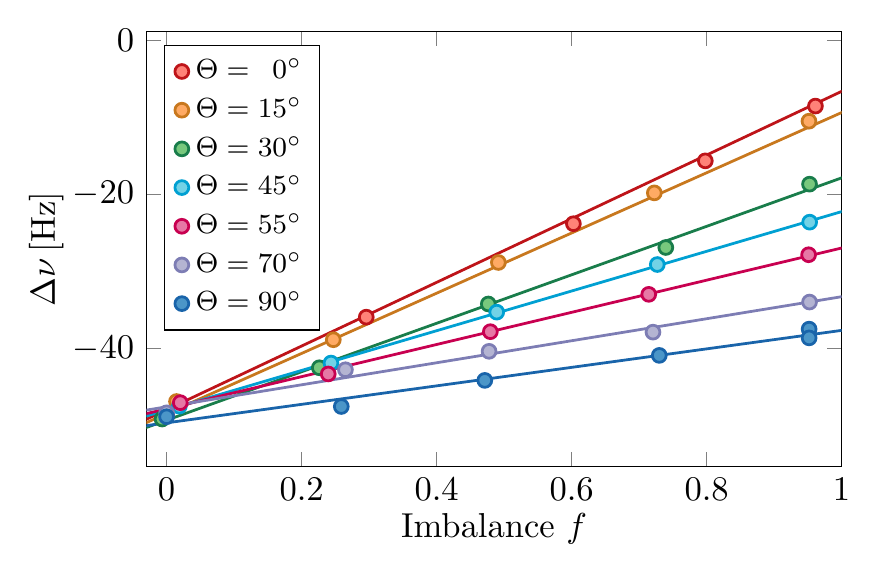}  
\caption{(a-b) Normalized Ramsey oscillations measured for ${\boldsymbol B}$ perpendicular ($\Theta=0^\circ$) or parallel ($\Theta=90^\circ$) to the atomic plane. For both cases, we show the transferred population as a function of detuning $\delta$ to the single-atom resonance. In each case the resonance is marked by a vertical dashed line. The circles (resp.\;squares) correspond to a balanced (resp.\;unbalanced) mixture $f=0$ (resp.\;$f\approx 0.95$).
Vertical error bars represent the standard deviation from the two measurements realized for each points.
(c) Variation of the frequency shift $\Delta \nu$ with the transferred fraction $f$. We restrict to positive imbalances, for which the population in $|2\rangle$ remains small enough to limit the role of two-body relaxation and spin-changing collisions.  For each angle, the solid line is a linear fit to the data. 
}\label{fig2}
\end{figure}

We now tackle the experimental observation of this modification of the inter-species scattering length in a quasi-2D Bose gas. The experimental setup was described in \cite{Ville17,SaintJalm19}. Basically, a cloud of  $^{87}$Rb atoms in state $|1\rangle$ is confined in a 2D box potential:  A ``hard-wall" potential provides a uniform in-plane confinement inside a 12\,\micron  radius disk, unless otherwise stated.  The vertical confinement can be approximated by a harmonic potential with frequency $\omega_{z}/2\pi$ = 4.4(1)\,kHz, corresponding to $\ell_z=160\,$nm. We operate in the weakly interacting regime characterized by the dimensionless coupling constant $\tilde g=\sqrt{8\pi}\,a_{11}/\ell_z=0.16(1)$, where $a_{11}$ is the $s$-wave scattering length for atoms in $|1\rangle$. The in-plane density of the cloud is $\bar n \approx 95 /\micron^2$ and we operate at the lowest achievable temperature in our setup $T < 30\,$nK. A $\approx$ 0.7 Gauss bias magnetic field ${\boldsymbol B}$ with tunable orientation is fixed during the experiment. 

Spectroscopy is performed thanks to a Ramsey sequence similar to \cite{Harber02}. Atoms initially in $|1\rangle$ are coupled to state $|2\rangle$ with a microwave field tuned around the hyperfine splitting of 6.8\,GHz. A first Ramsey pulse with a typical duration of a few tens of  \microsec\, creates a superposition of the two clock states with a tunable weight. After an ``interrogation time" $T_R=10\,$ms, a second identical Ramsey pulse is applied \footnote{The imbalance $f$ is tuned mostly by changing the pulse duration but for small pulses area it is more convenient to also decrease the Rabi frequency to avoid using very short microwave pulses.}. After this second pulse, we perform absorption imaging to determine the population in $|2\rangle$. We measure the variation of this population as a function of the frequency of the microwave field, see Figs.\,\ref{fig2}ab. We fit a sinusoidal function to the data, so as to determine the resonance frequency of the atomic cloud. All frequency measurements $\Delta \nu$ are reported with respect to reference measurements of the single-atom response that we perform on a dilute cloud. The typical dispersion of the measurement of this single-atom response is about 1\,Hz and provides an estimate of our uncertainty on the frequency measurements. We checked that the measured resonance frequencies are independent of $T_R$ in the range 5-20\,ms. Shorter delays lead to a lower accuracy on the frequency measurement. For longer delays, we observe demixing dynamics \cite{Timmermans98} between the two components and a modification of the resonance frequency.

In the following, we restrict to the case of strongly degenerate clouds \footnote{At non-zero temperature, quantum statistics of thermal bosons lead to multiply this shift by a factor which varies from 1 in the very degenerate regime to 2 for a thermal cloud.}  described in the mean-field approximation. Consider first the case of a uniform 3D gas. The resonant frequency $\Delta \nu$ can be computed by evaluating the difference of mean-field shifts for the two components \cite{Harber02},
\begin{eqnarray}
\Delta \nu=\frac{\hbar}{m} n&\, \left[ a_{22}-a_{11}+(2 a_{12}-a_{11}-a_{22})f\right].
\label{eq:shift}
\end{eqnarray}
Here the $a_{ij}$ are the inter- and intra-species scattering lengths, $n=n_1+n_2$ is  the total 3D density of the cloud where each component $i$ has a density $n_i$ after the first Ramsey pulse and $f={(n_1-n_2)}/{(n_1+n_2)}$ describes the population imbalance between the two states. 

It is interesting to discuss briefly two limiting cases of Eq.\,(\ref{eq:shift}). In the low transfer limit $f\approx1$, the first Ramsey pulse produces only a few atoms in state $|2\rangle$, imbedded in a bath of state $|1\rangle$ atoms. Interactions within pairs of state $|2\rangle$ atoms then play a negligible role, so that the shift $\Delta \nu$ does not depend on $a_{22}$. It is proportional to $(a_{12}-a_{11})$, hence sensitive to MDDI. In the balanced case $f=0$, the Ramsey sequence transforms a gas initially composed only of atoms in state $|1\rangle$ into a gas composed only of atoms in state $|2\rangle$. The energy balance between initial and final states then gives a contribution  $\Delta \nu \propto (a_{22}-a_{11})$, which is insensitive to MDDI. 

It is important to note that the validity of Eq.\,(\ref{eq:shift}) for a many-body system is not straightforward and requires some care \cite{Martin13,Fletcher17}. We discuss  in Ref.\,\cite{Zou20} the applicability of this approach to our experimental system, and show that it relies on the almost equality of the three relevant scattering lengths $a_{ij}$ of the problem. Note also that in our geometry, even if the gas is uniform in plane, the density distribution along $z$ is inhomogeneous and the spectroscopy measurement is thus sensitive to the integrated density $\bar n(x,y)=\int \mathrm{d}z\;n(x,y,z)$.

\begin{figure}[t!]
  \centering
  \includegraphics{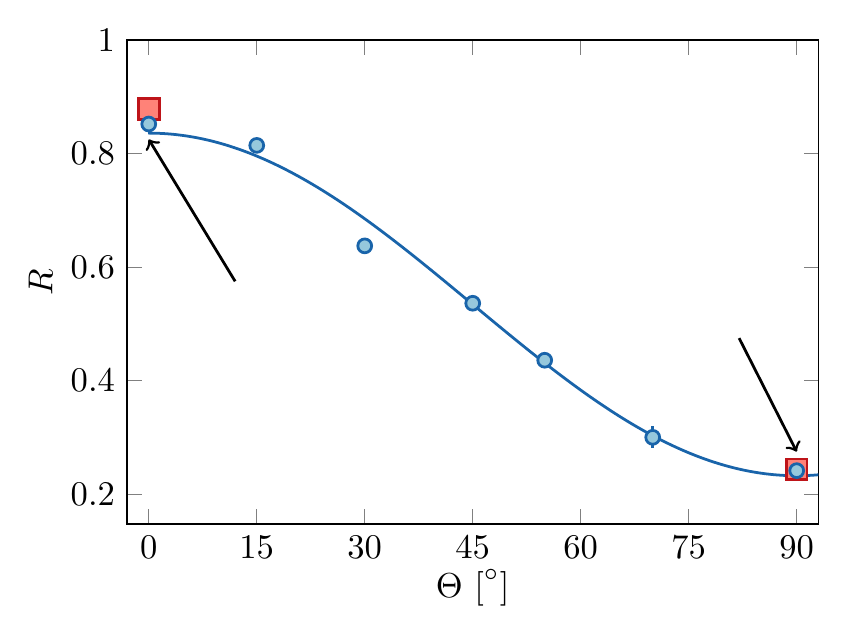}
  \vskip-100pt   \hskip-85pt
  \includegraphics[width=2.8cm]{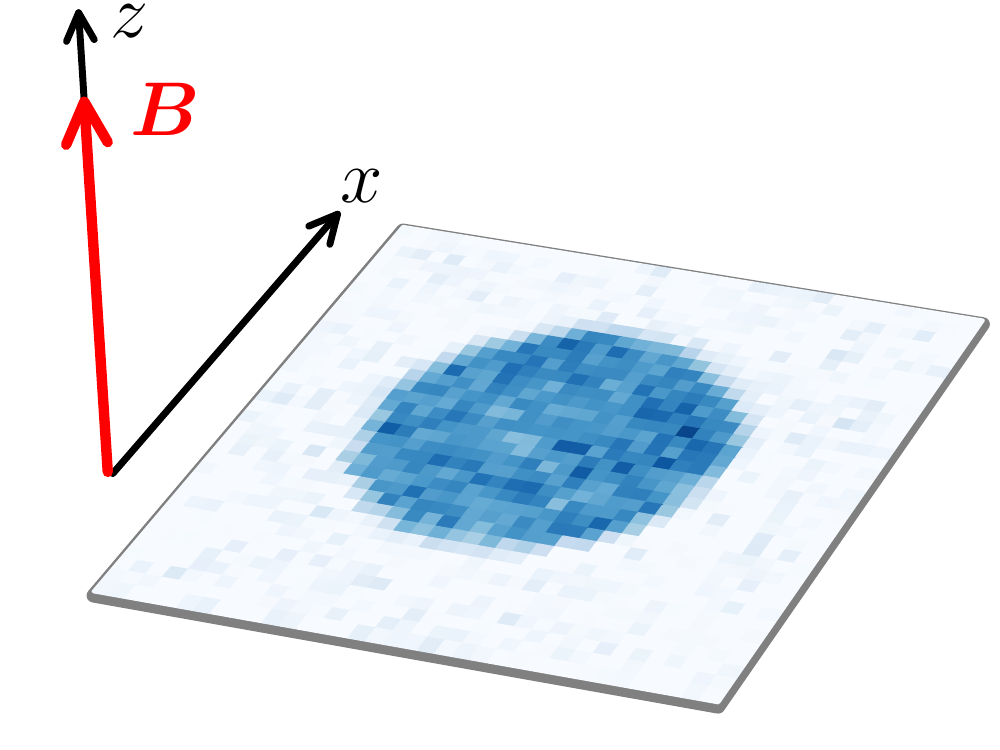}
  \vskip-108pt  \hskip144pt
  \includegraphics[width=2.8cm]{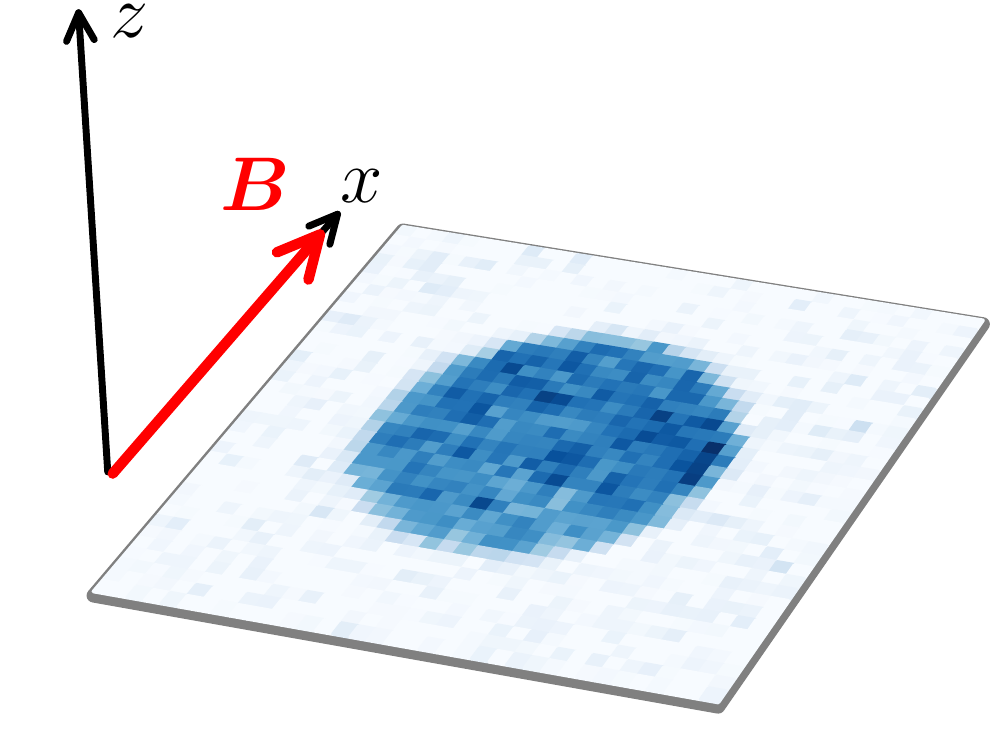}
  \vskip75pt 
  \caption{Variation of the ratio $R(\Theta)$ determined from the data of Fig.\,\ref{fig2}c with the magnetic field orientation $\Theta$. Blue circles (resp. red squares) correspond to the measurement at maximum density (resp. half density). The variation of this ratio is well fitted by a cosine variation compatible with the prediction for MDDI. The amplitude and offset of this variation allow one to determine accurately relative values of the scattering lengths. Vertical error bars represent the uncertainty obtained from the fitting procedure of the data in Fig.\ref{fig2}. The uncertainty on the determination of the angles is limited by the geometrical arrangement of the coils generating the field ${\boldsymbol B}$, estimated here at the level of 1$^\circ$.}\label{fig3}
\end{figure}

We now discuss the measurement of the frequency shift $\Delta \nu$ as a function of the imbalance $f$ for different orientations of the magnetic field with respect to the atomic plane, see Fig.\,\ref{fig2}c. For each orientation, we confirm the linear  behavior expected from Eq.\,\eqref{eq:shift}. The variation of the slope ${\rm d}\Delta \nu/{\rm d}f$ for different orientations reflects the expected modification of $a_{12}$ with $\Theta$ of Eq.\,\eqref{eq:da12}. More quantitatively, we fit a linear function to the data for each $\Theta$. The ratio of the slope to the intercept of this line is   $R(\Theta)={\left[a_{22}+a_{11}-2a_{12}(\Theta)\right]}/{(a_{22}-a_{11})}$. Interestingly, this ratio is independent of the density calibration and is thus a robust observable. 

The evolution of the measured ratio for different angles is shown in Fig.\,\ref{fig3}. For  $\Theta=0^\circ$ and $90^\circ$, we also show the ratio measured for a density approximately twice smaller than the one of Fig.\,\ref{fig2}. These two points overlap well with the main curve, which confirms the insensitivity of $R$ with respect to $\bar n$. We fit a sinusoidal variation $\Theta \mapsto \alpha+\beta \cos(2\Theta)$ to $R(\Theta)$ from which we extract $\alpha=0.53(1)$ and $\beta=0.30(1)$. We then determine $a_{22}-a_{11}=-3 a_{\rm dd}/\beta$ and $a_{12}^{(0)}-a_{11}=a_{\rm dd}{(3\alpha-3-\beta)}/{(2\beta)}$. Using $a_{\rm dd}=0.70\,a_0$, with $a_0$ the Bohr radius, we find  $a_{22}-a_{11}=-7.0(2)\, a_0$ and  $a_{12}^{(0)}-a_{11}=-2.0(1)\,a_0$.
 These results are in good agreement with the values predicted in \cite{Altin11},  $a_{11}=100.9\,a_0$, $a_{22}-a_{11}=-6.0\,a_0$ and  $a_{12}^{(0)}-a_{11}=-2.0\,a_0$.

\begin{figure}[t!]
  \centering
  \includegraphics{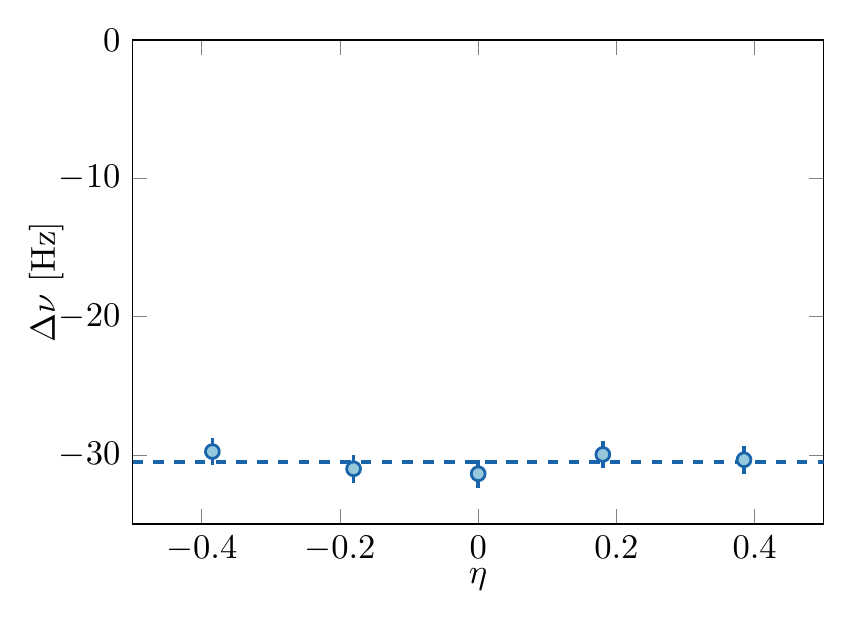}
  \vskip-85pt \hskip30pt
  \includegraphics[width=1cm]{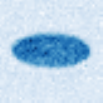}\hskip10pt
  \includegraphics[width=1cm]{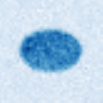}\hskip10pt
  \includegraphics[width=1cm]{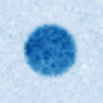}\hskip10pt
  \includegraphics[width=1cm]{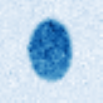}\hskip10pt
  \includegraphics[width=1cm]{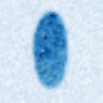}
  \vskip -41mm \hskip 30mm
  \includegraphics{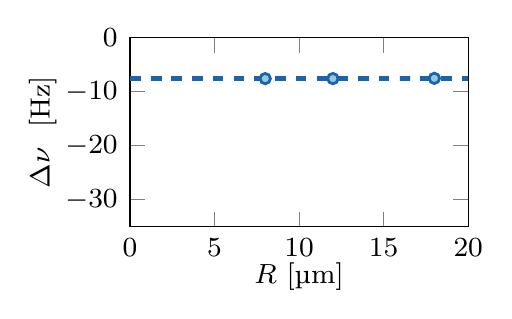}
  \vskip-58pt  \hskip135pt
  \includegraphics[width=0.7cm]{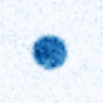}\hskip3pt
  \includegraphics[width=0.7cm]{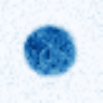}\hskip8pt
  \includegraphics[width=0.7cm]{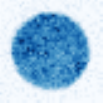}
  \vskip38mm
  \caption{Interaction shift $\Delta \nu$ as a function of the anisotropy parameter $\eta$. For a fixed density and an in-plane magnetic field, we vary the anisotropy of the elliptically-shaped 2D cloud. No dependence on the shape of the cloud is observed, in agreement with the expected isotropic character of MDDI in 2D when $R_{x,y}\gg \ell_z$. Vertical error bars represent the estimated 1\,Hz accuracy on the determination of the single-atom resonance frequency. Inset: interaction shift as a function of the size of the cloud, for ${\boldsymbol B}$ normal to the atom plane. }
  \label{fig4}
\end{figure}

All experiments described so far have been realized with a fixed disk geometry. As stated above, the description of the contribution of MDDI as a modification of the inter-species scattering length relies on the effective isotropy of the interaction in our 2D system. We investigate this issue by measuring the frequency shift of the clock transition for an in-plane magnetic field orientation ($\Theta=90^\circ$), which breaks the rotational symmetry of the system. We operate with a fixed density ($\bar n \approx 80 /\micron^2$)  and a varying elliptical shape. We choose a large imbalance $f\approx 0.95$ to have the highest sensitivity to possible modifications of $a_{12}$. We define an anisotropy parameter $\eta=(R_y-R_x)/(R_x+R_y)$ for the ratio of the lengths $R_x$ and $R_y$ of the two axes of the ellipse. We report in Fig.\,\ref{fig4} the measured shifts as a function of $\eta$ and confirm, within our experimental accuracy, the independence of the MDDI energy with respect to the cloud shape.
We have also investigated the influence of the size of the cloud on $\Delta \nu$ (inset of Fig.\,\ref{fig4}). Here we choose a disk-shaped cloud and a magnetic field perpendicular to the atomic plane. We observe no detectable change of $\Delta \nu$ when changing the disk radius from 8 to 18\,\micron\hspace{-2pt}, which confirms the absence of significant finite-size effects.

In conclusion, thanks to high resolution spectroscopy we revealed the non-negligible role of magnetic dipolar interactions between states with a zero average magnetic moment. We observed and explained the modification of the inter-species scattering length in a two-component cloud. Because of the smallness of MDDI for alkali-metal atoms, we did not observe any modification of the global shape of the cloud. This contrasts with the case of single-component highly-magnetic dipolar gases where the shape of a trapped gas has been modified with a static \cite{ODell04,Stuhler05,Lahaye07} or time-averaged-field \cite{Giovanazzi02,Tang18}. Nevertheless, the effect observed here provides a novel control on the dynamics of two-component gases. For example, the effective interaction parameter between two atoms in state $|2\rangle$ mediated by a bath of atoms in state $|1\rangle$ can be written as $\tilde g_{22}^{\rm eff}=\tilde g_{22}-\tilde g_{12}^2/\tilde g_{11}$, where $\tilde g_{ij}=\sqrt{8\pi } a_{ij}/\ell_z$ \cite{Pethick08}. With our parameters, we achieve a variation by a factor 7 of $\tilde g_{22}^{\rm eff}$, which will lead to important modifications of polaron dynamics. Similarly, it can be exploited to tune the miscibility of mixtures or the dynamics of spin textures. The distance to the critical point for miscibility, whose position is given by $\tilde g_{22}\tilde g_{11}=\tilde g_{12}^2$,  is also strongly sensitive to a variation of $\tilde g_{12}$. For instance,  the length scale of spin textures appearing in phase separation dynamics of a balanced mixture  will be modified, for our parameters, by a factor of almost 3 when $\Theta$ is switched from 0 to $90^\circ$ \cite{Timmermans98}. In addition, one can exploit the non-local character of MDDI by confining the atoms in a deep lattice at unit filling, where the exchange coupling evidenced here will implement the so-called quantum XX model \cite{Sachdev} without requiring any tunneling between lattice sites. The extreme sensitivity of the clock transition and its protection from magnetic perturbations will then provide a novel, precise tool to detect the various phases of matter predicted within this model.

\begin{acknowledgments}
This work is supported by ERC (Synergy UQUAM), QuantERA ERA-NET (NAQUAS project) and the ANR-18-CE30-0010 grant. We thank F.\;Pereira dos Santos, M. Zwierlein and P.\;Julienne for stimulating discussions. We acknowledge the contribution of R.\;Saint-Jalm at the early stage of the project.
\end{acknowledgments}

\bibliography{bib_clock}

\begin{thebibliography}{43}%
\makeatletter
\providecommand \@ifxundefined [1]{%
 \@ifx{#1\undefined}
}%
\providecommand \@ifnum [1]{%
 \ifnum #1\expandafter \@firstoftwo
 \else \expandafter \@secondoftwo
 \fi
}%
\providecommand \@ifx [1]{%
 \ifx #1\expandafter \@firstoftwo
 \else \expandafter \@secondoftwo
 \fi
}%
\providecommand \natexlab [1]{#1}%
\providecommand \enquote  [1]{``#1''}%
\providecommand \bibnamefont  [1]{#1}%
\providecommand \bibfnamefont [1]{#1}%
\providecommand \citenamefont [1]{#1}%
\providecommand \href@noop [0]{\@secondoftwo}%
\providecommand \href [0]{\begingroup \@sanitize@url \@href}%
\providecommand \@href[1]{\@@startlink{#1}\@@href}%
\providecommand \@@href[1]{\endgroup#1\@@endlink}%
\providecommand \@sanitize@url [0]{\catcode `\\12\catcode `\$12\catcode
  `\&12\catcode `\#12\catcode `\^12\catcode `\_12\catcode `\%12\relax}%
\providecommand \@@startlink[1]{}%
\providecommand \@@endlink[0]{}%
\providecommand \url  [0]{\begingroup\@sanitize@url \@url }%
\providecommand \@url [1]{\endgroup\@href {#1}{\urlprefix }}%
\providecommand \urlprefix  [0]{URL }%
\providecommand \Eprint [0]{\href }%
\providecommand \doibase [0]{http://dx.doi.org/}%
\providecommand \selectlanguage [0]{\@gobble}%
\providecommand \bibinfo  [0]{\@secondoftwo}%
\providecommand \bibfield  [0]{\@secondoftwo}%
\providecommand \translation [1]{[#1]}%
\providecommand \BibitemOpen [0]{}%
\providecommand \bibitemStop [0]{}%
\providecommand \bibitemNoStop [0]{.\EOS\space}%
\providecommand \EOS [0]{\spacefactor3000\relax}%
\providecommand \BibitemShut  [1]{\csname bibitem#1\endcsname}%
\let\auto@bib@innerbib\@empty
\bibitem [{\citenamefont {Chin}\ \emph {et~al.}(2010)\citenamefont {Chin},
  \citenamefont {Grimm}, \citenamefont {Julienne},\ and\ \citenamefont
  {Tiesinga}}]{Chin08}%
  \BibitemOpen
  \bibfield  {author} {\bibinfo {author} {\bibfnamefont {C.}~\bibnamefont
  {Chin}}, \bibinfo {author} {\bibfnamefont {R.}~\bibnamefont {Grimm}},
  \bibinfo {author} {\bibfnamefont {P.}~\bibnamefont {Julienne}}, \ and\
  \bibinfo {author} {\bibfnamefont {E.}~\bibnamefont {Tiesinga}},\ }\bibfield
  {title} {\enquote {\bibinfo {title} {Feshbach resonances in ultracold
  gases},}\ }\href {\doibase 10.1103/RevModPhys.82.1225} {\bibfield  {journal}
  {\bibinfo  {journal} {Rev. Mod. Phys.}\ }\textbf {\bibinfo {volume} {82}},\
  \bibinfo {pages} {1225} (\bibinfo {year} {2010})}\BibitemShut {NoStop}%
\bibitem [{\citenamefont {Bloch}\ \emph {et~al.}(2008)\citenamefont {Bloch},
  \citenamefont {Dalibard},\ and\ \citenamefont {Zwerger}}]{Bloch08}%
  \BibitemOpen
  \bibfield  {author} {\bibinfo {author} {\bibfnamefont {I.}~\bibnamefont
  {Bloch}}, \bibinfo {author} {\bibfnamefont {J.}~\bibnamefont {Dalibard}}, \
  and\ \bibinfo {author} {\bibfnamefont {W.}~\bibnamefont {Zwerger}},\
  }\bibfield  {title} {\enquote {\bibinfo {title} {Many-body physics with
  ultracold gases},}\ }\href {\doibase 10.1103/RevModPhys.80.885} {\bibfield
  {journal} {\bibinfo  {journal} {Rev. Mod. Phys.}\ }\textbf {\bibinfo {volume}
  {80}},\ \bibinfo {pages} {885} (\bibinfo {year} {2008})}\BibitemShut
  {NoStop}%
\bibitem [{\citenamefont {Baumann}\ \emph {et~al.}(2010)\citenamefont
  {Baumann}, \citenamefont {Guerlin}, \citenamefont {Brennecke},\ and\
  \citenamefont {Esslinger}}]{Baumann10}%
  \BibitemOpen
  \bibfield  {author} {\bibinfo {author} {\bibfnamefont {K.}~\bibnamefont
  {Baumann}}, \bibinfo {author} {\bibfnamefont {C.}~\bibnamefont {Guerlin}},
  \bibinfo {author} {\bibfnamefont {F.}~\bibnamefont {Brennecke}}, \ and\
  \bibinfo {author} {\bibfnamefont {T.}~\bibnamefont {Esslinger}},\ }\bibfield
  {title} {\enquote {\bibinfo {title} {Dicke quantum phase transition with a
  superfluid gas in an optical cavity},}\ }\href@noop {} {\bibfield  {journal}
  {\bibinfo  {journal} {Nature}\ }\textbf {\bibinfo {volume} {464}},\ \bibinfo
  {pages} {1301} (\bibinfo {year} {2010})}\BibitemShut {NoStop}%
\bibitem [{\citenamefont {L{\"o}w}\ \emph {et~al.}(2012)\citenamefont
  {L{\"o}w}, \citenamefont {Weimer}, \citenamefont {Nipper}, \citenamefont
  {Balewski}, \citenamefont {Butscher}, \citenamefont {B{\"u}chler},\ and\
  \citenamefont {Pfau}}]{Low12}%
  \BibitemOpen
  \bibfield  {author} {\bibinfo {author} {\bibfnamefont {R.}~\bibnamefont
  {L{\"o}w}}, \bibinfo {author} {\bibfnamefont {H.}~\bibnamefont {Weimer}},
  \bibinfo {author} {\bibfnamefont {J.}~\bibnamefont {Nipper}}, \bibinfo
  {author} {\bibfnamefont {J.B.}\ \bibnamefont {Balewski}}, \bibinfo {author}
  {\bibfnamefont {B.}~\bibnamefont {Butscher}}, \bibinfo {author}
  {\bibfnamefont {H.P.}\ \bibnamefont {B{\"u}chler}}, \ and\ \bibinfo {author}
  {\bibfnamefont {T.}~\bibnamefont {Pfau}},\ }\bibfield  {title} {\enquote
  {\bibinfo {title} {An experimental and theoretical guide to strongly
  interacting {R}ydberg gases},}\ }\href@noop {} {\bibfield  {journal}
  {\bibinfo  {journal} {J. Phys. B: At., Mol. Opt. Phys.}\ }\textbf {\bibinfo
  {volume} {45}},\ \bibinfo {pages} {113001} (\bibinfo {year}
  {2012})}\BibitemShut {NoStop}%
\bibitem [{\citenamefont {Lahaye}\ \emph {et~al.}(2009)\citenamefont {Lahaye},
  \citenamefont {Menotti}, \citenamefont {Santos}, \citenamefont {Lewenstein},\
  and\ \citenamefont {Pfau}}]{Lahaye09}%
  \BibitemOpen
  \bibfield  {author} {\bibinfo {author} {\bibfnamefont {T.}~\bibnamefont
  {Lahaye}}, \bibinfo {author} {\bibfnamefont {C.}~\bibnamefont {Menotti}},
  \bibinfo {author} {\bibfnamefont {L.}~\bibnamefont {Santos}}, \bibinfo
  {author} {\bibfnamefont {M.}~\bibnamefont {Lewenstein}}, \ and\ \bibinfo
  {author} {\bibfnamefont {T.}~\bibnamefont {Pfau}},\ }\bibfield  {title}
  {\enquote {\bibinfo {title} {The physics of dipolar bosonic quantum gases},}\
  }\href@noop {} {\bibfield  {journal} {\bibinfo  {journal} {Rep. Prog. Phys.}\
  }\textbf {\bibinfo {volume} {72}},\ \bibinfo {pages} {126401} (\bibinfo
  {year} {2009})}\BibitemShut {NoStop}%
\bibitem [{\citenamefont {Ferrier-Barbut}\ \emph {et~al.}(2016)\citenamefont
  {Ferrier-Barbut}, \citenamefont {Kadau}, \citenamefont {Schmitt},
  \citenamefont {Wenzel},\ and\ \citenamefont {Pfau}}]{Ferrier16}%
  \BibitemOpen
  \bibfield  {author} {\bibinfo {author} {\bibfnamefont {I.}~\bibnamefont
  {Ferrier-Barbut}}, \bibinfo {author} {\bibfnamefont {H.}~\bibnamefont
  {Kadau}}, \bibinfo {author} {\bibfnamefont {M.}~\bibnamefont {Schmitt}},
  \bibinfo {author} {\bibfnamefont {M.}~\bibnamefont {Wenzel}}, \ and\ \bibinfo
  {author} {\bibfnamefont {T.}~\bibnamefont {Pfau}},\ }\bibfield  {title}
  {\enquote {\bibinfo {title} {Observation of quantum droplets in a strongly
  dipolar {B}ose gas},}\ }\href {\doibase 10.1103/PhysRevLett.116.215301}
  {\bibfield  {journal} {\bibinfo  {journal} {Phys. Rev. Lett.}\ }\textbf
  {\bibinfo {volume} {116}},\ \bibinfo {pages} {215301} (\bibinfo {year}
  {2016})}\BibitemShut {NoStop}%
\bibitem [{\citenamefont {Chomaz}\ \emph {et~al.}(2018)\citenamefont {Chomaz},
  \citenamefont {van Bijnen}, \citenamefont {Petter}, \citenamefont {Faraoni},
  \citenamefont {Baier}, \citenamefont {Becher}, \citenamefont {Mark},
  \citenamefont {Waechtler}, \citenamefont {Santos},\ and\ \citenamefont
  {Ferlaino}}]{Chomaz18}%
  \BibitemOpen
  \bibfield  {author} {\bibinfo {author} {\bibfnamefont {L.}~\bibnamefont
  {Chomaz}}, \bibinfo {author} {\bibfnamefont {R.M.W.}\ \bibnamefont {van
  Bijnen}}, \bibinfo {author} {\bibfnamefont {D.}~\bibnamefont {Petter}},
  \bibinfo {author} {\bibfnamefont {G.}~\bibnamefont {Faraoni}}, \bibinfo
  {author} {\bibfnamefont {S.}~\bibnamefont {Baier}}, \bibinfo {author}
  {\bibfnamefont {J.H.}\ \bibnamefont {Becher}}, \bibinfo {author}
  {\bibfnamefont {M.J.}\ \bibnamefont {Mark}}, \bibinfo {author} {\bibfnamefont
  {F.}~\bibnamefont {Waechtler}}, \bibinfo {author} {\bibfnamefont
  {L.}~\bibnamefont {Santos}}, \ and\ \bibinfo {author} {\bibfnamefont
  {F.}~\bibnamefont {Ferlaino}},\ }\bibfield  {title} {\enquote {\bibinfo
  {title} {Observation of roton mode population in a dipolar quantum gas},}\
  }\href@noop {} {\bibfield  {journal} {\bibinfo  {journal} {Nat. Phys.}\
  }\textbf {\bibinfo {volume} {14}},\ \bibinfo {pages} {442} (\bibinfo {year}
  {2018})}\BibitemShut {NoStop}%
\bibitem [{\citenamefont {de~Paz}\ \emph {et~al.}(2013)\citenamefont {de~Paz},
  \citenamefont {Sharma}, \citenamefont {Chotia}, \citenamefont {Mar\'echal},
  \citenamefont {Huckans}, \citenamefont {Pedri}, \citenamefont {Santos},
  \citenamefont {Gorceix}, \citenamefont {Vernac},\ and\ \citenamefont
  {Laburthe-Tolra}}]{Paz13}%
  \BibitemOpen
  \bibfield  {author} {\bibinfo {author} {\bibfnamefont {A.}~\bibnamefont
  {de~Paz}}, \bibinfo {author} {\bibfnamefont {A.}~\bibnamefont {Sharma}},
  \bibinfo {author} {\bibfnamefont {A.}~\bibnamefont {Chotia}}, \bibinfo
  {author} {\bibfnamefont {E.}~\bibnamefont {Mar\'echal}}, \bibinfo {author}
  {\bibfnamefont {J.H.}\ \bibnamefont {Huckans}}, \bibinfo {author}
  {\bibfnamefont {P.}~\bibnamefont {Pedri}}, \bibinfo {author} {\bibfnamefont
  {L.}~\bibnamefont {Santos}}, \bibinfo {author} {\bibfnamefont
  {O.}~\bibnamefont {Gorceix}}, \bibinfo {author} {\bibfnamefont
  {L.}~\bibnamefont {Vernac}}, \ and\ \bibinfo {author} {\bibfnamefont
  {B.}~\bibnamefont {Laburthe-Tolra}},\ }\bibfield  {title} {\enquote {\bibinfo
  {title} {Nonequilibrium quantum magnetism in a dipolar lattice gas},}\ }\href
  {\doibase 10.1103/PhysRevLett.111.185305} {\bibfield  {journal} {\bibinfo
  {journal} {Phys. Rev. Lett.}\ }\textbf {\bibinfo {volume} {111}},\ \bibinfo
  {pages} {185305} (\bibinfo {year} {2013})}\BibitemShut {NoStop}%
\bibitem [{\citenamefont {Baier}\ \emph {et~al.}(2016)\citenamefont {Baier},
  \citenamefont {Mark}, \citenamefont {Petter}, \citenamefont {Aikawa},
  \citenamefont {Chomaz}, \citenamefont {Cai}, \citenamefont {Baranov},
  \citenamefont {Zoller},\ and\ \citenamefont {Ferlaino}}]{Baier16}%
  \BibitemOpen
  \bibfield  {author} {\bibinfo {author} {\bibfnamefont {S.}~\bibnamefont
  {Baier}}, \bibinfo {author} {\bibfnamefont {M.J.}\ \bibnamefont {Mark}},
  \bibinfo {author} {\bibfnamefont {D.}~\bibnamefont {Petter}}, \bibinfo
  {author} {\bibfnamefont {K.}~\bibnamefont {Aikawa}}, \bibinfo {author}
  {\bibfnamefont {L.}~\bibnamefont {Chomaz}}, \bibinfo {author} {\bibfnamefont
  {Z.}~\bibnamefont {Cai}}, \bibinfo {author} {\bibfnamefont {M.}~\bibnamefont
  {Baranov}}, \bibinfo {author} {\bibfnamefont {P.}~\bibnamefont {Zoller}}, \
  and\ \bibinfo {author} {\bibfnamefont {F.}~\bibnamefont {Ferlaino}},\
  }\bibfield  {title} {\enquote {\bibinfo {title} {Extended {Bose-Hubbard
  }models with ultracold magnetic atoms},}\ }\href {\doibase
  10.1126/science.aac9812} {\bibfield  {journal} {\bibinfo  {journal}
  {Science}\ }\textbf {\bibinfo {volume} {352}},\ \bibinfo {pages} {201}
  (\bibinfo {year} {2016})}\BibitemShut {NoStop}%
\bibitem [{\citenamefont {Lepoutre}\ \emph {et~al.}(2019)\citenamefont
  {Lepoutre}, \citenamefont {Schachenmayer}, \citenamefont {Gabardos},
  \citenamefont {Zhu}, \citenamefont {Naylor}, \citenamefont {Mar{\'e}chal},
  \citenamefont {Gorceix}, \citenamefont {Rey}, \citenamefont {Vernac},\ and\
  \citenamefont {Laburthe-Tolra}}]{lepoutre2019out}%
  \BibitemOpen
  \bibfield  {author} {\bibinfo {author} {\bibfnamefont {S.}~\bibnamefont
  {Lepoutre}}, \bibinfo {author} {\bibfnamefont {J.}~\bibnamefont
  {Schachenmayer}}, \bibinfo {author} {\bibfnamefont {L.}~\bibnamefont
  {Gabardos}}, \bibinfo {author} {\bibfnamefont {B.}~\bibnamefont {Zhu}},
  \bibinfo {author} {\bibfnamefont {B.}~\bibnamefont {Naylor}}, \bibinfo
  {author} {\bibfnamefont {E.}~\bibnamefont {Mar{\'e}chal}}, \bibinfo {author}
  {\bibfnamefont {O.}~\bibnamefont {Gorceix}}, \bibinfo {author} {\bibfnamefont
  {A.M.}\ \bibnamefont {Rey}}, \bibinfo {author} {\bibfnamefont
  {L.}~\bibnamefont {Vernac}}, \ and\ \bibinfo {author} {\bibfnamefont
  {B.}~\bibnamefont {Laburthe-Tolra}},\ }\bibfield  {title} {\enquote {\bibinfo
  {title} {Out-of-equilibrium quantum magnetism and thermalization in a spin-3
  many-body dipolar lattice system},}\ }\href@noop {} {\bibfield  {journal}
  {\bibinfo  {journal} {Nat. Commun.}\ }\textbf {\bibinfo {volume} {10}},\
  \bibinfo {pages} {1} (\bibinfo {year} {2019})}\BibitemShut {NoStop}%
\bibitem [{\citenamefont {Giovanazzi}\ \emph {et~al.}(2002)\citenamefont
  {Giovanazzi}, \citenamefont {G\"orlitz},\ and\ \citenamefont
  {Pfau}}]{Giovanazzi02}%
  \BibitemOpen
  \bibfield  {author} {\bibinfo {author} {\bibfnamefont {S.}~\bibnamefont
  {Giovanazzi}}, \bibinfo {author} {\bibfnamefont {A.}~\bibnamefont
  {G\"orlitz}}, \ and\ \bibinfo {author} {\bibfnamefont {T.}~\bibnamefont
  {Pfau}},\ }\bibfield  {title} {\enquote {\bibinfo {title} {Tuning the dipolar
  interaction in quantum gases},}\ }\href {\doibase
  10.1103/PhysRevLett.89.130401} {\bibfield  {journal} {\bibinfo  {journal}
  {Phys. Rev. Lett.}\ }\textbf {\bibinfo {volume} {89}},\ \bibinfo {pages}
  {130401} (\bibinfo {year} {2002})}\BibitemShut {NoStop}%
\bibitem [{\citenamefont {Fattori}\ \emph {et~al.}(2008)\citenamefont
  {Fattori}, \citenamefont {Roati}, \citenamefont {Deissler}, \citenamefont
  {D'Errico}, \citenamefont {Zaccanti}, \citenamefont {Jona-Lasinio},
  \citenamefont {Santos}, \citenamefont {Inguscio},\ and\ \citenamefont
  {Modugno}}]{Fattori08}%
  \BibitemOpen
  \bibfield  {author} {\bibinfo {author} {\bibfnamefont {M.}~\bibnamefont
  {Fattori}}, \bibinfo {author} {\bibfnamefont {G.}~\bibnamefont {Roati}},
  \bibinfo {author} {\bibfnamefont {B.}~\bibnamefont {Deissler}}, \bibinfo
  {author} {\bibfnamefont {C.}~\bibnamefont {D'Errico}}, \bibinfo {author}
  {\bibfnamefont {M.}~\bibnamefont {Zaccanti}}, \bibinfo {author}
  {\bibfnamefont {M.}~\bibnamefont {Jona-Lasinio}}, \bibinfo {author}
  {\bibfnamefont {L.}~\bibnamefont {Santos}}, \bibinfo {author} {\bibfnamefont
  {M.}~\bibnamefont {Inguscio}}, \ and\ \bibinfo {author} {\bibfnamefont
  {G.}~\bibnamefont {Modugno}},\ }\bibfield  {title} {\enquote {\bibinfo
  {title} {Magnetic dipolar interaction in a {Bose-Einstein} condensate atomic
  interferometer},}\ }\href {\doibase 10.1103/PhysRevLett.101.190405}
  {\bibfield  {journal} {\bibinfo  {journal} {Phys. Rev. Lett.}\ }\textbf
  {\bibinfo {volume} {101}},\ \bibinfo {pages} {190405} (\bibinfo {year}
  {2008})}\BibitemShut {NoStop}%
\bibitem [{\citenamefont {Pollack}\ \emph {et~al.}(2009)\citenamefont
  {Pollack}, \citenamefont {Dries}, \citenamefont {Junker}, \citenamefont
  {Chen}, \citenamefont {Corcovilos},\ and\ \citenamefont {Hulet}}]{Pollack09}%
  \BibitemOpen
  \bibfield  {author} {\bibinfo {author} {\bibfnamefont {S.E.}\ \bibnamefont
  {Pollack}}, \bibinfo {author} {\bibfnamefont {D.}~\bibnamefont {Dries}},
  \bibinfo {author} {\bibfnamefont {M.}~\bibnamefont {Junker}}, \bibinfo
  {author} {\bibfnamefont {Y.~P.}\ \bibnamefont {Chen}}, \bibinfo {author}
  {\bibfnamefont {T.A.}\ \bibnamefont {Corcovilos}}, \ and\ \bibinfo {author}
  {\bibfnamefont {R.G.}\ \bibnamefont {Hulet}},\ }\bibfield  {title} {\enquote
  {\bibinfo {title} {Extreme tunability of interactions in a $^{7}\mathrm{Li}$
  {Bose-Einstein} condensate},}\ }\href {\doibase
  10.1103/PhysRevLett.102.090402} {\bibfield  {journal} {\bibinfo  {journal}
  {Phys. Rev. Lett.}\ }\textbf {\bibinfo {volume} {102}},\ \bibinfo {pages}
  {090402} (\bibinfo {year} {2009})}\BibitemShut {NoStop}%
\bibitem [{\citenamefont {Stamper-Kurn}\ and\ \citenamefont
  {Ueda}(2013)}]{StamperKurn13}%
  \BibitemOpen
  \bibfield  {author} {\bibinfo {author} {\bibfnamefont {D.M.}\ \bibnamefont
  {Stamper-Kurn}}\ and\ \bibinfo {author} {\bibfnamefont {M.}~\bibnamefont
  {Ueda}},\ }\bibfield  {title} {\enquote {\bibinfo {title} {Spinor {B}ose
  gases: {S}ymmetries, magnetism, and quantum dynamics},}\ }\href {\doibase
  10.1103/RevModPhys.85.1191} {\bibfield  {journal} {\bibinfo  {journal} {Rev.
  Mod. Phys.}\ }\textbf {\bibinfo {volume} {85}},\ \bibinfo {pages}
  {1191--1244} (\bibinfo {year} {2013})}\BibitemShut {NoStop}%
\bibitem [{\citenamefont {Yi}\ \emph {et~al.}(2004)\citenamefont {Yi},
  \citenamefont {You},\ and\ \citenamefont {Pu}}]{Yi04}%
  \BibitemOpen
  \bibfield  {author} {\bibinfo {author} {\bibfnamefont {S.}~\bibnamefont
  {Yi}}, \bibinfo {author} {\bibfnamefont {L.}~\bibnamefont {You}}, \ and\
  \bibinfo {author} {\bibfnamefont {H.}~\bibnamefont {Pu}},\ }\bibfield
  {title} {\enquote {\bibinfo {title} {Quantum phases of dipolar spinor
  condensates},}\ }\href {\doibase 10.1103/PhysRevLett.93.040403} {\bibfield
  {journal} {\bibinfo  {journal} {Phys. Rev. Lett.}\ }\textbf {\bibinfo
  {volume} {93}},\ \bibinfo {pages} {040403} (\bibinfo {year}
  {2004})}\BibitemShut {NoStop}%
\bibitem [{\citenamefont {Vengalattore}\ \emph {et~al.}(2008)\citenamefont
  {Vengalattore}, \citenamefont {Leslie}, \citenamefont {Guzman},\ and\
  \citenamefont {Stamper-Kurn}}]{Vengalattore08}%
  \BibitemOpen
  \bibfield  {author} {\bibinfo {author} {\bibfnamefont {M.}~\bibnamefont
  {Vengalattore}}, \bibinfo {author} {\bibfnamefont {S.R.}\ \bibnamefont
  {Leslie}}, \bibinfo {author} {\bibfnamefont {J.}~\bibnamefont {Guzman}}, \
  and\ \bibinfo {author} {\bibfnamefont {D.M.}\ \bibnamefont {Stamper-Kurn}},\
  }\bibfield  {title} {\enquote {\bibinfo {title} {Spontaneously modulated spin
  textures in a dipolar spinor {Bose-Einstein} condensate},}\ }\href {\doibase
  10.1103/PhysRevLett.100.170403} {\bibfield  {journal} {\bibinfo  {journal}
  {Phys. Rev. Lett.}\ }\textbf {\bibinfo {volume} {100}},\ \bibinfo {pages}
  {170403} (\bibinfo {year} {2008})}\BibitemShut {NoStop}%
\bibitem [{\citenamefont {Eto}\ \emph {et~al.}(2014)\citenamefont {Eto},
  \citenamefont {Saito},\ and\ \citenamefont {Hirano}}]{Eto14}%
  \BibitemOpen
  \bibfield  {author} {\bibinfo {author} {\bibfnamefont {Y.}~\bibnamefont
  {Eto}}, \bibinfo {author} {\bibfnamefont {H.}~\bibnamefont {Saito}}, \ and\
  \bibinfo {author} {\bibfnamefont {T.}~\bibnamefont {Hirano}},\ }\bibfield
  {title} {\enquote {\bibinfo {title} {Observation of dipole-induced spin
  texture in an $^{87}\mathrm{Rb}$ {Bose-Einstein} condensate},}\ }\href
  {\doibase 10.1103/PhysRevLett.112.185301} {\bibfield  {journal} {\bibinfo
  {journal} {Phys. Rev. Lett.}\ }\textbf {\bibinfo {volume} {112}},\ \bibinfo
  {pages} {185301} (\bibinfo {year} {2014})}\BibitemShut {NoStop}%
\bibitem [{\citenamefont {Marti}\ \emph {et~al.}(2014)\citenamefont {Marti},
  \citenamefont {MacRae}, \citenamefont {Olf}, \citenamefont {Lourette},
  \citenamefont {Fang},\ and\ \citenamefont {Stamper-Kurn}}]{Marti14}%
  \BibitemOpen
  \bibfield  {author} {\bibinfo {author} {\bibfnamefont {G.E.}\ \bibnamefont
  {Marti}}, \bibinfo {author} {\bibfnamefont {A.}~\bibnamefont {MacRae}},
  \bibinfo {author} {\bibfnamefont {R.}~\bibnamefont {Olf}}, \bibinfo {author}
  {\bibfnamefont {S.}~\bibnamefont {Lourette}}, \bibinfo {author}
  {\bibfnamefont {F.}~\bibnamefont {Fang}}, \ and\ \bibinfo {author}
  {\bibfnamefont {D.M.}\ \bibnamefont {Stamper-Kurn}},\ }\bibfield  {title}
  {\enquote {\bibinfo {title} {Coherent magnon optics in a ferromagnetic spinor
  {Bose-Einstein} condensate},}\ }\href {\doibase
  10.1103/PhysRevLett.113.155302} {\bibfield  {journal} {\bibinfo  {journal}
  {Phys. Rev. Lett.}\ }\textbf {\bibinfo {volume} {113}},\ \bibinfo {pages}
  {155302} (\bibinfo {year} {2014})}\BibitemShut {NoStop}%
\bibitem [{\citenamefont {Yan}\ \emph {et~al.}(2013)\citenamefont {Yan},
  \citenamefont {Moses}, \citenamefont {Gadway}, \citenamefont {Covey},
  \citenamefont {Hazzard}, \citenamefont {Rey}, \citenamefont {Jin},\ and\
  \citenamefont {Ye}}]{yan2013observation}%
  \BibitemOpen
  \bibfield  {author} {\bibinfo {author} {\bibfnamefont {B.}~\bibnamefont
  {Yan}}, \bibinfo {author} {\bibfnamefont {S.A.}\ \bibnamefont {Moses}},
  \bibinfo {author} {\bibfnamefont {B.}~\bibnamefont {Gadway}}, \bibinfo
  {author} {\bibfnamefont {J.P.}\ \bibnamefont {Covey}}, \bibinfo {author}
  {\bibfnamefont {K.R.A.}\ \bibnamefont {Hazzard}}, \bibinfo {author}
  {\bibfnamefont {A.M.}\ \bibnamefont {Rey}}, \bibinfo {author} {\bibfnamefont
  {D.S.}\ \bibnamefont {Jin}}, \ and\ \bibinfo {author} {\bibfnamefont
  {J.}~\bibnamefont {Ye}},\ }\bibfield  {title} {\enquote {\bibinfo {title}
  {Observation of dipolar spin-exchange interactions with lattice-confined
  polar molecules},}\ }\href@noop {} {\bibfield  {journal} {\bibinfo  {journal}
  {Nature}\ }\textbf {\bibinfo {volume} {501}},\ \bibinfo {pages} {521}
  (\bibinfo {year} {2013})}\BibitemShut {NoStop}%
\bibitem [{\citenamefont {de~L\'es\'eleuc}\ \emph {et~al.}(2017)\citenamefont
  {de~L\'es\'eleuc}, \citenamefont {Barredo}, \citenamefont {Lienhard},
  \citenamefont {Browaeys},\ and\ \citenamefont
  {Lahaye}}]{Leseleuc:2017_PhysRevLett.119.053202}%
  \BibitemOpen
  \bibfield  {author} {\bibinfo {author} {\bibfnamefont {S.}~\bibnamefont
  {de~L\'es\'eleuc}}, \bibinfo {author} {\bibfnamefont {D.}~\bibnamefont
  {Barredo}}, \bibinfo {author} {\bibfnamefont {V.}~\bibnamefont {Lienhard}},
  \bibinfo {author} {\bibfnamefont {A.}~\bibnamefont {Browaeys}}, \ and\
  \bibinfo {author} {\bibfnamefont {T.}~\bibnamefont {Lahaye}},\ }\bibfield
  {title} {\enquote {\bibinfo {title} {Optical control of the resonant
  dipole-dipole interaction between {R}ydberg atoms},}\ }\href {\doibase
  10.1103/PhysRevLett.119.053202} {\bibfield  {journal} {\bibinfo  {journal}
  {Phys. Rev. Lett.}\ }\textbf {\bibinfo {volume} {119}},\ \bibinfo {pages}
  {053202} (\bibinfo {year} {2017})}\BibitemShut {NoStop}%
\bibitem [{Note1()}]{Note1}%
  \BibitemOpen
  \bibinfo {note} {For all experiments reported here, the fraction of atoms in
  any other spin state remains below our detection sensitivity of
  1\%.}\BibitemShut {Stop}%
\bibitem [{REF()}]{REFSM}%
  \BibitemOpen
  \href@noop {} {\enquote {\bibinfo {title} {For details, see {S}upplemental
  {M}aterials},}\ }\BibitemShut {NoStop}%
\bibitem [{\citenamefont {Fischer}(2006)}]{Fischer06}%
  \BibitemOpen
  \bibfield  {author} {\bibinfo {author} {\bibfnamefont {U.R.}\ \bibnamefont
  {Fischer}},\ }\bibfield  {title} {\enquote {\bibinfo {title} {Stability of
  quasi-two-dimensional {Bose-Einstein} condensates with dominant dipole-dipole
  interactions},}\ }\href {\doibase 10.1103/PhysRevA.73.031602} {\bibfield
  {journal} {\bibinfo  {journal} {Phys. Rev. A}\ }\textbf {\bibinfo {volume}
  {73}},\ \bibinfo {pages} {031602(R)} (\bibinfo {year} {2006})}\BibitemShut
  {NoStop}%
\bibitem [{\citenamefont {Fedorov}\ \emph {et~al.}(2014)\citenamefont
  {Fedorov}, \citenamefont {Kurbakov}, \citenamefont {Shchadilova},\ and\
  \citenamefont {Lozovik}}]{Fedorov14}%
  \BibitemOpen
  \bibfield  {author} {\bibinfo {author} {\bibfnamefont {A.K.}\ \bibnamefont
  {Fedorov}}, \bibinfo {author} {\bibfnamefont {I.L.}\ \bibnamefont
  {Kurbakov}}, \bibinfo {author} {\bibfnamefont {Y.E.}\ \bibnamefont
  {Shchadilova}}, \ and\ \bibinfo {author} {\bibfnamefont {Yu.E.}\ \bibnamefont
  {Lozovik}},\ }\bibfield  {title} {\enquote {\bibinfo {title} {Two-dimensional
  {B}ose gas of tilted dipoles: Roton instability and condensate depletion},}\
  }\href {\doibase 10.1103/PhysRevA.90.043616} {\bibfield  {journal} {\bibinfo
  {journal} {Phys. Rev. A}\ }\textbf {\bibinfo {volume} {90}},\ \bibinfo
  {pages} {043616} (\bibinfo {year} {2014})}\BibitemShut {NoStop}%
\bibitem [{\citenamefont {Mishra}\ and\ \citenamefont {Nath}(2016)}]{Mishra16}%
  \BibitemOpen
  \bibfield  {author} {\bibinfo {author} {\bibfnamefont {C.}~\bibnamefont
  {Mishra}}\ and\ \bibinfo {author} {\bibfnamefont {R.}~\bibnamefont {Nath}},\
  }\bibfield  {title} {\enquote {\bibinfo {title} {Dipolar condensates with
  tilted dipoles in a pancake-shaped confinement},}\ }\href {\doibase
  10.1103/PhysRevA.94.033633} {\bibfield  {journal} {\bibinfo  {journal} {Phys.
  Rev. A}\ }\textbf {\bibinfo {volume} {94}},\ \bibinfo {pages} {033633}
  (\bibinfo {year} {2016})}\BibitemShut {NoStop}%
\bibitem [{\citenamefont {Yi}\ and\ \citenamefont {You}(2000)}]{Yi00}%
  \BibitemOpen
  \bibfield  {author} {\bibinfo {author} {\bibfnamefont {S.}~\bibnamefont
  {Yi}}\ and\ \bibinfo {author} {\bibfnamefont {L.}~\bibnamefont {You}},\
  }\bibfield  {title} {\enquote {\bibinfo {title} {Trapped atomic condensates
  with anisotropic interactions},}\ }\href {\doibase
  10.1103/PhysRevA.61.041604} {\bibfield  {journal} {\bibinfo  {journal} {Phys.
  Rev. A}\ }\textbf {\bibinfo {volume} {61}},\ \bibinfo {pages} {041604}
  (\bibinfo {year} {2000})}\BibitemShut {NoStop}%
\bibitem [{Note2()}]{Note2}%
  \BibitemOpen
  \bibinfo {note} {We use the definition of Ref. \cite {Lahaye09}. Other
  definitions with a different numerical factor are found in the literature,
  see for instance, \cite {Bortolotti06,Baranov12}}\BibitemShut {NoStop}%
\bibitem [{\citenamefont {Ville}\ \emph {et~al.}(2017)\citenamefont {Ville},
  \citenamefont {Bienaim\'e}, \citenamefont {Saint-Jalm}, \citenamefont
  {Corman}, \citenamefont {Aidelsburger}, \citenamefont {Chomaz}, \citenamefont
  {Kleinlein}, \citenamefont {Perconte}, \citenamefont {Nascimb\`ene},
  \citenamefont {Dalibard},\ and\ \citenamefont {Beugnon}}]{Ville17}%
  \BibitemOpen
  \bibfield  {author} {\bibinfo {author} {\bibfnamefont {J.~L.}\ \bibnamefont
  {Ville}}, \bibinfo {author} {\bibfnamefont {T.}~\bibnamefont {Bienaim\'e}},
  \bibinfo {author} {\bibfnamefont {R.}~\bibnamefont {Saint-Jalm}}, \bibinfo
  {author} {\bibfnamefont {L.}~\bibnamefont {Corman}}, \bibinfo {author}
  {\bibfnamefont {M.}~\bibnamefont {Aidelsburger}}, \bibinfo {author}
  {\bibfnamefont {L.}~\bibnamefont {Chomaz}}, \bibinfo {author} {\bibfnamefont
  {K.}~\bibnamefont {Kleinlein}}, \bibinfo {author} {\bibfnamefont
  {D.}~\bibnamefont {Perconte}}, \bibinfo {author} {\bibfnamefont
  {S.}~\bibnamefont {Nascimb\`ene}}, \bibinfo {author} {\bibfnamefont
  {J.}~\bibnamefont {Dalibard}}, \ and\ \bibinfo {author} {\bibfnamefont
  {J.}~\bibnamefont {Beugnon}},\ }\bibfield  {title} {\enquote {\bibinfo
  {title} {Loading and compression of a single two-dimensional {B}ose gas in an
  optical accordion},}\ }\href {\doibase 10.1103/PhysRevA.95.013632} {\bibfield
   {journal} {\bibinfo  {journal} {Phys. Rev. A}\ }\textbf {\bibinfo {volume}
  {95}},\ \bibinfo {pages} {013632} (\bibinfo {year} {2017})}\BibitemShut
  {NoStop}%
\bibitem [{\citenamefont {Saint-Jalm}\ \emph {et~al.}(2019)\citenamefont
  {Saint-Jalm}, \citenamefont {Castilho}, \citenamefont {Le~Cerf},
  \citenamefont {Bakkali-Hassani}, \citenamefont {Ville}, \citenamefont
  {Nascimbene}, \citenamefont {Beugnon},\ and\ \citenamefont
  {Dalibard}}]{SaintJalm19}%
  \BibitemOpen
  \bibfield  {author} {\bibinfo {author} {\bibfnamefont {R.}~\bibnamefont
  {Saint-Jalm}}, \bibinfo {author} {\bibfnamefont {P.~C.~M.}\ \bibnamefont
  {Castilho}}, \bibinfo {author} {\bibfnamefont {\'E.}\ \bibnamefont
  {Le~Cerf}}, \bibinfo {author} {\bibfnamefont {B.}~\bibnamefont
  {Bakkali-Hassani}}, \bibinfo {author} {\bibfnamefont {J.-L.}\ \bibnamefont
  {Ville}}, \bibinfo {author} {\bibfnamefont {S.}~\bibnamefont {Nascimbene}},
  \bibinfo {author} {\bibfnamefont {J.}~\bibnamefont {Beugnon}}, \ and\
  \bibinfo {author} {\bibfnamefont {J.}~\bibnamefont {Dalibard}},\ }\bibfield
  {title} {\enquote {\bibinfo {title} {Dynamical symmetry and breathers in a
  two-dimensional {B}ose gas},}\ }\href {\doibase 10.1103/PhysRevX.9.021035}
  {\bibfield  {journal} {\bibinfo  {journal} {Phys. Rev. X}\ }\textbf {\bibinfo
  {volume} {9}},\ \bibinfo {pages} {021035} (\bibinfo {year}
  {2019})}\BibitemShut {NoStop}%
\bibitem [{\citenamefont {Harber}\ \emph {et~al.}(2002)\citenamefont {Harber},
  \citenamefont {Lewandowski}, \citenamefont {McGuirk},\ and\ \citenamefont
  {Cornell}}]{Harber02}%
  \BibitemOpen
  \bibfield  {author} {\bibinfo {author} {\bibfnamefont {D.M.}\ \bibnamefont
  {Harber}}, \bibinfo {author} {\bibfnamefont {H.J.}\ \bibnamefont
  {Lewandowski}}, \bibinfo {author} {\bibfnamefont {J.M.}\ \bibnamefont
  {McGuirk}}, \ and\ \bibinfo {author} {\bibfnamefont {E.A.}\ \bibnamefont
  {Cornell}},\ }\bibfield  {title} {\enquote {\bibinfo {title} {Effect of cold
  collisions on spin coherence and resonance shifts in a magnetically trapped
  ultracold gas},}\ }\href {\doibase 10.1103/PhysRevA.66.053616} {\bibfield
  {journal} {\bibinfo  {journal} {Phys. Rev. A}\ }\textbf {\bibinfo {volume}
  {66}},\ \bibinfo {pages} {053616} (\bibinfo {year} {2002})}\BibitemShut
  {NoStop}%
\bibitem [{Note3()}]{Note3}%
  \BibitemOpen
  \bibinfo {note} {The imbalance $f$ is tuned mostly by changing the pulse
  duration but for small pulses area it is more convenient to also decrease the
  Rabi frequency to avoid using very short microwave pulses.}\BibitemShut
  {Stop}%
\bibitem [{\citenamefont {Timmermans}(1998)}]{Timmermans98}%
  \BibitemOpen
  \bibfield  {author} {\bibinfo {author} {\bibfnamefont {E.}~\bibnamefont
  {Timmermans}},\ }\bibfield  {title} {\enquote {\bibinfo {title} {Phase
  separation of {Bose-E}instein condensates},}\ }\href {\doibase
  10.1103/PhysRevLett.81.5718} {\bibfield  {journal} {\bibinfo  {journal}
  {Phys. Rev. Lett.}\ }\textbf {\bibinfo {volume} {81}},\ \bibinfo {pages}
  {5718} (\bibinfo {year} {1998})}\BibitemShut {NoStop}%
\bibitem [{Note4()}]{Note4}%
  \BibitemOpen
  \bibinfo {note} {At non-zero temperature, quantum statistics of thermal
  bosons lead to multiply this shift by a factor which varies from 1 in the
  very degenerate regime to 2 for a thermal cloud.}\BibitemShut {Stop}%
\bibitem [{\citenamefont {Martin}\ \emph {et~al.}(2013)\citenamefont {Martin},
  \citenamefont {Bishof}, \citenamefont {Swallows}, \citenamefont {Zhang},
  \citenamefont {Benko}, \citenamefont {von Stecher}, \citenamefont {Gorshkov},
  \citenamefont {Rey},\ and\ \citenamefont {Ye}}]{Martin13}%
  \BibitemOpen
  \bibfield  {author} {\bibinfo {author} {\bibfnamefont {M.J.}\ \bibnamefont
  {Martin}}, \bibinfo {author} {\bibfnamefont {M.}~\bibnamefont {Bishof}},
  \bibinfo {author} {\bibfnamefont {M.D.}\ \bibnamefont {Swallows}}, \bibinfo
  {author} {\bibfnamefont {X.}~\bibnamefont {Zhang}}, \bibinfo {author}
  {\bibfnamefont {C.}~\bibnamefont {Benko}}, \bibinfo {author} {\bibfnamefont
  {J.}~\bibnamefont {von Stecher}}, \bibinfo {author} {\bibfnamefont {A.V.}\
  \bibnamefont {Gorshkov}}, \bibinfo {author} {\bibfnamefont {A.M.}\
  \bibnamefont {Rey}}, \ and\ \bibinfo {author} {\bibfnamefont
  {J.}~\bibnamefont {Ye}},\ }\bibfield  {title} {\enquote {\bibinfo {title} {A
  quantum many-body spin system in an optical lattice clock},}\ }\href
  {\doibase 10.1126/science.1236929} {\bibfield  {journal} {\bibinfo  {journal}
  {Science}\ }\textbf {\bibinfo {volume} {341}},\ \bibinfo {pages} {632}
  (\bibinfo {year} {2013})}\BibitemShut {NoStop}%
\bibitem [{\citenamefont {Fletcher}\ \emph {et~al.}(2017)\citenamefont
  {Fletcher}, \citenamefont {Lopes}, \citenamefont {Man}, \citenamefont
  {Navon}, \citenamefont {Smith}, \citenamefont {Zwierlein},\ and\
  \citenamefont {Hadzibabic}}]{Fletcher17}%
  \BibitemOpen
  \bibfield  {author} {\bibinfo {author} {\bibfnamefont {R.J.}\ \bibnamefont
  {Fletcher}}, \bibinfo {author} {\bibfnamefont {R.}~\bibnamefont {Lopes}},
  \bibinfo {author} {\bibfnamefont {J.}~\bibnamefont {Man}}, \bibinfo {author}
  {\bibfnamefont {N.}~\bibnamefont {Navon}}, \bibinfo {author} {\bibfnamefont
  {R.P.}\ \bibnamefont {Smith}}, \bibinfo {author} {\bibfnamefont {M.W.}\
  \bibnamefont {Zwierlein}}, \ and\ \bibinfo {author} {\bibfnamefont
  {Z.}~\bibnamefont {Hadzibabic}},\ }\bibfield  {title} {\enquote {\bibinfo
  {title} {Two-and three-body contacts in the unitary {B}ose gas},}\
  }\href@noop {} {\bibfield  {journal} {\bibinfo  {journal} {Science}\ }\textbf
  {\bibinfo {volume} {355}},\ \bibinfo {pages} {377} (\bibinfo {year}
  {2017})}\BibitemShut {NoStop}%
\bibitem [{\citenamefont {Zou}\ \emph {et~al.}(2020)\citenamefont {Zou},
  \citenamefont {Bakkali-Hassani}, \citenamefont {Maury}, \citenamefont
  {Le~Cerf}, \citenamefont {Nascimbene}, \citenamefont {Dalibard},\ and\
  \citenamefont {Beugnon}}]{Zou20}%
  \BibitemOpen
  \bibfield  {author} {\bibinfo {author} {\bibfnamefont {Y.-Q.}\ \bibnamefont
  {Zou}}, \bibinfo {author} {\bibfnamefont {B.}~\bibnamefont
  {Bakkali-Hassani}}, \bibinfo {author} {\bibfnamefont {C.}~\bibnamefont
  {Maury}}, \bibinfo {author} {\bibfnamefont {\'E.}\ \bibnamefont {Le~Cerf}},
  \bibinfo {author} {\bibfnamefont {S.}~\bibnamefont {Nascimbene}}, \bibinfo
  {author} {\bibfnamefont {J.}~\bibnamefont {Dalibard}}, \ and\ \bibinfo
  {author} {\bibfnamefont {J.}~\bibnamefont {Beugnon}},\ }\bibfield  {title}
  {\enquote {\bibinfo {title} {{T}an's two-body contact across the superfluid
  transition of a planar {B}ose gas},}\ }\href@noop {} {\bibfield  {journal}
  {\bibinfo  {journal} {arXiv:2007.12385}\ } (\bibinfo {year}
  {2020})}\BibitemShut {NoStop}%
\bibitem [{\citenamefont {Altin}\ \emph {et~al.}(2011)\citenamefont {Altin},
  \citenamefont {McDonald}, \citenamefont {D\"oring}, \citenamefont {Debs},
  \citenamefont {Barter}, \citenamefont {Close}, \citenamefont {Robins},
  \citenamefont {Haine}, \citenamefont {Hanna},\ and\ \citenamefont
  {Anderson}}]{Altin11}%
  \BibitemOpen
  \bibfield  {author} {\bibinfo {author} {\bibfnamefont {P.A.}\ \bibnamefont
  {Altin}}, \bibinfo {author} {\bibfnamefont {G.}~\bibnamefont {McDonald}},
  \bibinfo {author} {\bibfnamefont {D.}~\bibnamefont {D\"oring}}, \bibinfo
  {author} {\bibfnamefont {J.E.}\ \bibnamefont {Debs}}, \bibinfo {author}
  {\bibfnamefont {T.H.}\ \bibnamefont {Barter}}, \bibinfo {author}
  {\bibfnamefont {J.D.}\ \bibnamefont {Close}}, \bibinfo {author}
  {\bibfnamefont {N.P.}\ \bibnamefont {Robins}}, \bibinfo {author}
  {\bibfnamefont {S.A.}\ \bibnamefont {Haine}}, \bibinfo {author}
  {\bibfnamefont {T.M.}\ \bibnamefont {Hanna}}, \ and\ \bibinfo {author}
  {\bibfnamefont {R.P.}\ \bibnamefont {Anderson}},\ }\bibfield  {title}
  {\enquote {\bibinfo {title} {Optically trapped atom interferometry using the
  clock transition of large $^{87}${Rb} {B}ose--{Einstein} condensates},}\
  }\href {\doibase 10.1088/1367-2630/13/6/065020} {\bibfield  {journal}
  {\bibinfo  {journal} {New J. Phys.}\ }\textbf {\bibinfo {volume} {13}},\
  \bibinfo {pages} {065020} (\bibinfo {year} {2011})}\BibitemShut {NoStop}%
\bibitem [{\citenamefont {O'Dell}\ \emph {et~al.}(2004)\citenamefont {O'Dell},
  \citenamefont {Giovanazzi},\ and\ \citenamefont {Eberlein}}]{ODell04}%
  \BibitemOpen
  \bibfield  {author} {\bibinfo {author} {\bibfnamefont {D.H.~J.}\ \bibnamefont
  {O'Dell}}, \bibinfo {author} {\bibfnamefont {S.}~\bibnamefont {Giovanazzi}},
  \ and\ \bibinfo {author} {\bibfnamefont {C.}~\bibnamefont {Eberlein}},\
  }\bibfield  {title} {\enquote {\bibinfo {title} {Exact hydrodynamics of a
  trapped dipolar {Bose-Einstein} condensate},}\ }\href {\doibase
  10.1103/PhysRevLett.92.250401} {\bibfield  {journal} {\bibinfo  {journal}
  {Phys. Rev. Lett.}\ }\textbf {\bibinfo {volume} {92}},\ \bibinfo {pages}
  {250401} (\bibinfo {year} {2004})}\BibitemShut {NoStop}%
\bibitem [{\citenamefont {Stuhler}\ \emph {et~al.}(2005)\citenamefont
  {Stuhler}, \citenamefont {Griesmaier}, \citenamefont {Koch}, \citenamefont
  {Fattori}, \citenamefont {Pfau}, \citenamefont {Giovanazzi}, \citenamefont
  {Pedri},\ and\ \citenamefont {Santos}}]{Stuhler05}%
  \BibitemOpen
  \bibfield  {author} {\bibinfo {author} {\bibfnamefont {J.}~\bibnamefont
  {Stuhler}}, \bibinfo {author} {\bibfnamefont {A.}~\bibnamefont {Griesmaier}},
  \bibinfo {author} {\bibfnamefont {T.}~\bibnamefont {Koch}}, \bibinfo {author}
  {\bibfnamefont {M.}~\bibnamefont {Fattori}}, \bibinfo {author} {\bibfnamefont
  {T.}~\bibnamefont {Pfau}}, \bibinfo {author} {\bibfnamefont {S.}~\bibnamefont
  {Giovanazzi}}, \bibinfo {author} {\bibfnamefont {P.}~\bibnamefont {Pedri}}, \
  and\ \bibinfo {author} {\bibfnamefont {L.}~\bibnamefont {Santos}},\
  }\bibfield  {title} {\enquote {\bibinfo {title} {Observation of dipole-dipole
  interaction in a degenerate quantum gas},}\ }\href {\doibase
  10.1103/PhysRevLett.95.150406} {\bibfield  {journal} {\bibinfo  {journal}
  {Phys. Rev. Lett.}\ }\textbf {\bibinfo {volume} {95}},\ \bibinfo {pages}
  {150406} (\bibinfo {year} {2005})}\BibitemShut {NoStop}%
\bibitem [{\citenamefont {Lahaye}\ \emph {et~al.}(2007)\citenamefont {Lahaye},
  \citenamefont {Koch}, \citenamefont {Fr{\"o}hlich}, \citenamefont {Fattori},
  \citenamefont {Metz}, \citenamefont {Griesmaier}, \citenamefont
  {Giovanazzi},\ and\ \citenamefont {Pfau}}]{Lahaye07}%
  \BibitemOpen
  \bibfield  {author} {\bibinfo {author} {\bibfnamefont {T.}~\bibnamefont
  {Lahaye}}, \bibinfo {author} {\bibfnamefont {T.}~\bibnamefont {Koch}},
  \bibinfo {author} {\bibfnamefont {B.}~\bibnamefont {Fr{\"o}hlich}}, \bibinfo
  {author} {\bibfnamefont {M.}~\bibnamefont {Fattori}}, \bibinfo {author}
  {\bibfnamefont {J.}~\bibnamefont {Metz}}, \bibinfo {author} {\bibfnamefont
  {A.}~\bibnamefont {Griesmaier}}, \bibinfo {author} {\bibfnamefont
  {S.}~\bibnamefont {Giovanazzi}}, \ and\ \bibinfo {author} {\bibfnamefont
  {T.}~\bibnamefont {Pfau}},\ }\bibfield  {title} {\enquote {\bibinfo {title}
  {Strong dipolar effects in a quantum ferrofluid},}\ }\href@noop {} {\bibfield
   {journal} {\bibinfo  {journal} {Nature}\ }\textbf {\bibinfo {volume}
  {448}},\ \bibinfo {pages} {672} (\bibinfo {year} {2007})}\BibitemShut
  {NoStop}%
\bibitem [{\citenamefont {Tang}\ \emph {et~al.}(2018)\citenamefont {Tang},
  \citenamefont {Kao}, \citenamefont {Li},\ and\ \citenamefont {Lev}}]{Tang18}%
  \BibitemOpen
  \bibfield  {author} {\bibinfo {author} {\bibfnamefont {Y.}~\bibnamefont
  {Tang}}, \bibinfo {author} {\bibfnamefont {W.}~\bibnamefont {Kao}}, \bibinfo
  {author} {\bibfnamefont {K.Y.}\ \bibnamefont {Li}}, \ and\ \bibinfo {author}
  {\bibfnamefont {B.L.}\ \bibnamefont {Lev}},\ }\bibfield  {title} {\enquote
  {\bibinfo {title} {Tuning the dipole-dipole interaction in a quantum gas with
  a rotating magnetic field},}\ }\href {\doibase
  10.1103/PhysRevLett.120.230401} {\bibfield  {journal} {\bibinfo  {journal}
  {Phys. Rev. Lett.}\ }\textbf {\bibinfo {volume} {120}},\ \bibinfo {pages}
  {230401} (\bibinfo {year} {2018})}\BibitemShut {NoStop}%
\bibitem [{\citenamefont {Pethick}\ and\ \citenamefont
  {Smith}(2008)}]{Pethick08}%
  \BibitemOpen
  \bibfield  {author} {\bibinfo {author} {\bibfnamefont {C.J.}\ \bibnamefont
  {Pethick}}\ and\ \bibinfo {author} {\bibfnamefont {H.}~\bibnamefont
  {Smith}},\ }\href@noop {} {\emph {\bibinfo {title} {{Bose--Einstein
  }condensation in dilute gases}}}\ (\bibinfo  {publisher} {Cambridge
  {U}niversity {P}ress},\ \bibinfo {year} {2008})\BibitemShut {NoStop}%
\bibitem [{\citenamefont {Sachdev}(2006)}]{Sachdev}%
  \BibitemOpen
  \bibfield  {author} {\bibinfo {author} {\bibfnamefont {S.}~\bibnamefont
  {Sachdev}},\ }\href@noop {} {\emph {\bibinfo {title} {Quantum Phase
  Transitions}}}\ (\bibinfo  {publisher} {Cambridge University Press},\
  \bibinfo {year} {2006})\BibitemShut {NoStop}%
\end{thebibliography}%

\newpage

\section{SUPPLEMENTAL MATERIAL}

\subsection{Restriction of the dipole-dipole interaction to the clock state manifold
}
In this section, we evaluate the action of the magnetic dipole-dipole interaction inside the two-level manifold relevant for the clock transition. First, using the general expression for the coupling between a spin $1/2$ (here the outer electron) and a spin $i$ (here the $^{87}$Rb nucleus with $i=3/2$), we obtain the decomposition of the clock states on the basis $|s_Z,i_Z\rangle$:
\begin{eqnarray}
|1\rangle &\equiv& |F=1,m_Z=0\rangle \nonumber\\
&=& \frac{1}{\sqrt 2}\,\left( -|-\frac{1}{2};+\frac{1}{2}\rangle \ +\ |+\frac{1}{2};-\frac{1}{2}\rangle\right)
\label{eq:clock_state1}
\end{eqnarray}
and
\begin{eqnarray}
|2\rangle &\equiv& |F=2,m=0\rangle \nonumber \\
&=& \frac{1}{\sqrt 2}\,\left( |-\frac{1}{2};+\frac{1}{2}\rangle \ +\ |+\frac{1}{2};-\frac{1}{2}\rangle \right).
\label{eq:clock_state2}
\end{eqnarray}

The magnetic interaction operator for two electronic spins $\hat {\boldsymbol{s}}_A$ and $\hat {\boldsymbol{s}}_B$ with magnetic moments $\boldsymbol{m}_{A,B}=2\mu_B \boldsymbol{s}_{A,B}$  is given by
\begin{eqnarray}
\hat V_{\rm dd}(r,\boldsymbol{u})=\frac{\mu_0 \mu_B^2}{\pi r^3 } [\hat {\boldsymbol{s}}_A \cdot \hat {\boldsymbol{s}}_B-3 (\hat {\boldsymbol{s}}_A \cdot \boldsymbol{u})(\hat {\boldsymbol{s}}_B  \cdot \boldsymbol{u})],
\label{eq:VDD_SM}
\end{eqnarray}
where $r$ is the distance between the two dipoles and $\boldsymbol{u}$ is the unit vector connecting them. We calculate the matrix elements of this operator in the basis $\{|11\rangle, |12\rangle, |21\rangle, |22\rangle \}$, restricting to elastic interactions which are the only relevant ones for the experimental time scale. This leaves us with four different matrix elements to compute: $V_{1111}$, $V_{2222}$, $V_{1212}=V_{2121}$ and $V_{1221}=V_{2112}$, where $V_{ijkl}=\langle kl|\hat V_{\rm dd}| ij\rangle$. The calculation in the basis (\ref{eq:clock_state1},\ref{eq:clock_state2}) 
 leads to 
 \begin{equation}
\hat s_Z|1\rangle=\frac{1}{2}|2\rangle, \qquad  \hat s_Z|2\rangle=\frac{1}{2}|1\rangle
\end{equation}
The operators $\hat s_X$ and $\hat s_Y$ couple states with different $m_F$ values and the associated matrix elements  $V_{ijkl}$ inside the clock state manifold are  zero.
The magnetic interaction operator in Eq.\,\eqref{eq:VDD_SM} thus simplifies to 
\begin{equation}
\hat V_{\rm dd}(r,\theta)=\frac{\mu_0 \mu_B^2}{\pi r^3 }\,(1-3 \cos^2\theta)\; \hat {s}_{Z,A} \, \hat {s}_{Z,B}.
\end{equation}
We deduce that among the four matrix elements mentioned above, only 
\begin{equation}
V_{1221}=V_{2112}=\frac{\mu_0 \mu_B^2}{4\pi r^3 }\,(1-3 \cos^2\theta)
\end{equation}
is non-zero, where $\theta$ is the angle between $\boldsymbol{u}$ and the quantization axis. This shows that MDDI do not modify the interactions between atoms in the same state $|1\rangle$ or $|2\rangle$, but induce a non-local, angle-dependent, exchange interaction. The second-quantized Hamiltonian of the MDDI for the clock transition is thus:
\begin{align}
\hat H_{\rm dd}^{(1,2)}&=\frac{\mu_0 \mu_B^2}{4\pi}\iint \mathrm{d}^3r_A \;\mathrm{d}^3r_B \ \frac{1-3 \cos^2\theta}{|\boldsymbol r_A-\boldsymbol r_B|^3} \nonumber\\ 
& \times\;\hat \Psi_2^\dagger(\boldsymbol{r}_A)\, \hat \Psi_1^\dagger(\boldsymbol{r}_B)\, \hat \Psi_2(\boldsymbol{r}_B)\, \hat \Psi_1(\boldsymbol{r}_A),
\label{eq:Hamiltonian_SM}
\end{align}
where the $\hat \Psi_i(\boldsymbol{r}_\alpha)$ are the field operators annihilating a particle in state $|i\rangle$ at position $\boldsymbol{r}_\alpha$.

\subsection{Calculation of $\delta a_{12}$}
We consider a gas with a density distribution $n(x,y,z)=\rho(x,y) e^{-z^2/\ell_z^2}/(\ell_z \sqrt{\pi})$ subject to a magnetic field $\boldsymbol{B}=B(\cos\Theta \, \boldsymbol{u}_z+\sin\Theta\, \boldsymbol{u}_x)$ which defines the quantization axis $Z$ for the spin states. We compute the mean-field energy associated to magnetic dipole-dipole interactions following Ref.\,\cite{Fischer06}. In Fourier space, we express the mean-field energy as
\begin{eqnarray}
\langle \hat H_{\rm dd} \rangle=\frac{1}{2} \frac{1}{(2\pi)^3} \int \mathrm{d}^3k \; \tilde n(\boldsymbol{k})\; \tilde V_{\rm dd}(\boldsymbol{k})\; \tilde n(-\boldsymbol{k})\label{eq:Hdd}
\end{eqnarray}
where $ \tilde V_{\rm dd}(\boldsymbol{k})=3\mu_0 \mu_B^2[\cos^2\alpha-1/3]$ is the Fourier transform of the dipole-dipole interaction with $\alpha$ the angle of the wavevector $\boldsymbol{k}$ with respect to $\boldsymbol{B}$. The Fourier transform of the density distribution is given by $\tilde n(\boldsymbol{k})=e^{k_z^2 \ell_z^2/4} \tilde \rho(k_x,k_y)$, where $\tilde \rho(k_x,k_y)$ is the Fourier transform of $\rho(x,y)$. Introducing $k=|\boldsymbol{k}|$, we have $\cos(\alpha)=(k_z\cos\Theta+ k_x \sin\Theta)/k$ and we get
\begin{widetext}
\begin{eqnarray}
\langle \hat H_{\rm dd} \rangle=\frac{\mu_0 \mu_B^2}{3 \sqrt{2\pi}\ell_z} \left\{\frac{1}{(2\pi)^2} \int \mathrm{d}^2k \; \tilde \rho(k_x,k_y) \; \tilde \rho(-k_x,-k_y) \left[(3\cos^2 \Theta-1)+\left(\frac{k_x^2}{k_\perp^2} \sin^2\Theta-\cos^2 \Theta \right) \mathcal{F}(k_\perp \ell_z)\right]\right\},\label{eq:Hdd2}
\end{eqnarray}
\end{widetext}
where we have introduced $k_\perp=\sqrt{k_x^2+k_y^2}$ and $\mathcal{F}(k_\perp \ell_z)=3 \sqrt{\pi/2} k_\perp \ell_z e^{k_\perp^2 \ell_z^2/2}\, \mathrm{erfc}(k_\perp \ell_z^2/\sqrt{2})$. For a uniform system, $\rho(k_x,k_y)=\delta(k_x)\delta(k_y)$ is the product of two Dirac delta functions and we recover Eq.(3) of the main text. Consider now the case of spins aligned along the $x$ axis corresponding to $\Theta=\pi/2$, as in Fig.\,4 of the main text. For a cloud shape with typical length scales larger than $\ell_z$, the influence of the shape of the cloud via the integration over $k_x$ and $k_y$ scales with $\mathcal{F}(k_\perp \ell_z) \sim k_\perp \ell_z $, which is a small parameter in the 2D case considered here. Thus, the mean-field shift is expected to be independent of the in-plane geometry of the cloud.

\end{document}